\documentclass[english,aps,twocolumn,prmaterials,groupedaddress,superscriptaddress,longbibliography]{revtex4-2}
\usepackage{graphicx,epsfig,units}
\usepackage{xcolor} 
\usepackage{soul} 
\usepackage{amsmath,amsfonts,mathrsfs,amsbsy,bm,babel}
\usepackage{changes}
\usepackage{natbib}

\definecolor{imcolor}{rgb}{0.5,0.,0.5}				

\definecolor{ohcolor}{rgb}{0.,0.5,0.5}

\begin{document}
\author{O. Heinonen}
\affiliation{Materials Science Division, Argonne National Laboratory, Lemont, Illinois 60439, USA}
\email{heinonen@anl.gov}
\author{R. A. Heinonen}
\affiliation{Department of Physics and INFN, University of Rome "Tor Vergata", Via della Ricerca Scientifica 1, 00133 Rome, Italy}
\author{H. Park}
\affiliation{Materials Science Division, Argonne National Laboratory, Lemont, Illinois 60439, USA}
\affiliation{Department of Physics, University of Illinois, Chicago, Illinois 60607, USA}
\title{Magnetic ground states of a model for (TM)Nb$_3$S$_6$, TM=Co, Fe, Ni}
\begin{abstract}
The transition-metal intercalated dichalcogenide CoNb$_3$S$_6$ is a triangular  antiferromagnet that has recently been shown to exhibit a large anomalous Hall effect (AHE) below the N\'eel temperature, even though the response to an external field is very small. This suggests that there is an interesting magnetic structure that interacts with the electronic structure to yield the AHE, as collinear antiferromagnets cannot exhibit a nonzero AHE. We propose a model for magnetic transition-metal intercalated dichalcogenides and examine its ground state as function of interaction parameters. The model exhibits transitions between planar spin spirals, non-planar spin spirals, and a particular non-coplanar so-called $3q$ state. This latter state must exhibit a nonzero AHE, while the spin spirals do not. 
\end{abstract}
\date{\today}
\maketitle
\section{Introduction}\label{sec:intro}

Ferromagnetic (FM) and antiferromagnetic (AFM) systems can be frustrated when all interactions cannot be simultaneously minimized. For FMs, frustration involves more than near-neighbor interactions. The frustration can in FMs lead to a number of ordered states with complex orders, such as spiral states and skyrmion crystals\cite{muhlbauer2009,nagaosa2010,leonov2015,zhang2017}. In contrast, AFMs are in a sense easier to frustrate because in certain lattices, frustration is a geometrical property and near-neighbor interactions alone will lead to frustration and complex orders\cite{moessner1998}. 
Classic examples of a frustrated AFM are the triangular Ising AFM, the ground state of which was solved in two dimensions (2D) by Wannier in 1944\cite{wannier,collins1997review}, or the 2D triangular XY and Heisenberg AFMs with N\'eel ground states in which the three spins on an elementary triangular plaquette are $120^\circ$ degrees apart. More generally, triangular AFMs can also exhibit a number of different collinear and noncollinear states\cite{liu2016}. Another class of frustrated AFMs are the Kagome AFMs\cite{reimers1993order,zhitomirsky2008octupolar,gvozdikova2011magnetic,grison2020}, such as Mn$_3$Ge. Antiferromagnets have recently become the focus of intense interest in the connection of topological materials and their magnetotransport properties. It turns out that there exist quite a few examples of materials that are Kagome antiferromagnets or ferrimagnets with linked magnetic and topological properties, for example Mn$_3$Ge and  Mn$_3$Sn\cite{chen2014anomalousMn3Ir,nakatsuji2015large,nayak2016large,zhang2017strong,ikhlas2017large,higo2018large,kimata2019magnetic,chen2020antichiral} as well as others, such as (RE)Mn$_6$Sn$_6$ with RE (Rare Earth)\cite{yin2020quantum,ghimire2020competing,asaba2020anomalous,ma2021rare}, and MnBi$_2$Te$_4$\cite{eremeev2017competing,otrokov2017highly,otrokov2019prediction,li2019intrinsic,deng2020quantum}. In addition to nearest-neighbor interactions that are ubiquitous in AFMs, if the crystal is not centrosymmetric, a chiral Dzyaloshinskii-Moriya interaction (DMI) is allowed. The direction of the DMI vector ${\mathbf d}_{ij}$ that couples spins at sites $i$ and $j$ in the same plane depends on the in-plane symmetry. For example, in Kagome AFMs the DMI vector is along the crystallographic $c$ axis, perpendicular to the Kagome plane. 

 A phenomenon that connects magnetotransport to topology of the electronic structure is the anomalous Hall effect (AHE). Modern theories directly relate the AHE to the Berry phase of the electronic bands in the first Brillouin zone (BZ)\cite{chang1995berry,chang1996berry,sundaram1999wave,jungwirth2002,nagaosa2010,xiao2010berry}. There are also direct connections between the real-space magnetic structure of AFMs and the AHE\cite{taguchi2001spin,martin2008itinerant,zhang2020real}. A real-space magnetic texture with a finite chirality can give rise to a fictitious magnetic field that, in turn, produces a Hall effect\cite{machida2010time,kato_stability_2010,solenov2012}. The chirality is defined as $\chi=\epsilon_{123}{\mathbf S}_1\cdot\left[ {\mathbf S}_2\times {\mathbf S}_3\right] $, where ${\mathbf S}_i$, $i=1,2,3$ are three spins on an elementary triangular plaquette for the case of triangular or Kagome systems, and $\epsilon_{ijk}$ is the Levi-Civita symbol. This is the same concept that, in the continuum limit, gives rise to a topological magnetic field and a topological Hall effect in magnetic skyrmions\cite{neubauer2009topological}. In the presence of spin-orbit coupling, the relation between real-space spin texture and Berry curvature becomes complicated. For example, gapless collinear AFMs cannot exhibit an AHE\cite{shindou2001orbital,surgers2014large,ghimire2018large}. Coplanar Kagome or triangular AFMs such as Mn$_3$Ge or PdCrO$_2$ can exhibit an AHE\cite{chen2014anomalousMn3Ir,nakatsuji2015Mn3Sn,takatsu2010} only in the presence of spin-orbit coupling or a small net moment that break certain symmetries\cite{chen2014anomalousMn3Ir}, while non-coplanar AFMs with non-zero chirality can exhibit a nonzero AHE\cite{taguchi2001spin,martin2008itinerant,zhang2020real}. In general,  
 if the system is invariant under the combination of time reversal ${\mathcal T}$ and a lattice translation ${\mathcal R}$, the Berry phase is zero, and the chirality is also zero. More generally, if the system is invariant under the combination of ${\mathcal T}$ and ${\mathcal O}$, where ${\mathcal O}$ is any unitary symmetry operator, the Berry phase is zero. Conversely, if the system is not invariant under ${\mathcal TR}$, the Berry phase and the chirality can both be nonzero. Therefore, a non-zero chirality for a non-coplanar system is an indication that there can be a non-zero Berry phase and a nonzero AHE.

\begin{figure}
    \centering
    \includegraphics[width=\columnwidth]{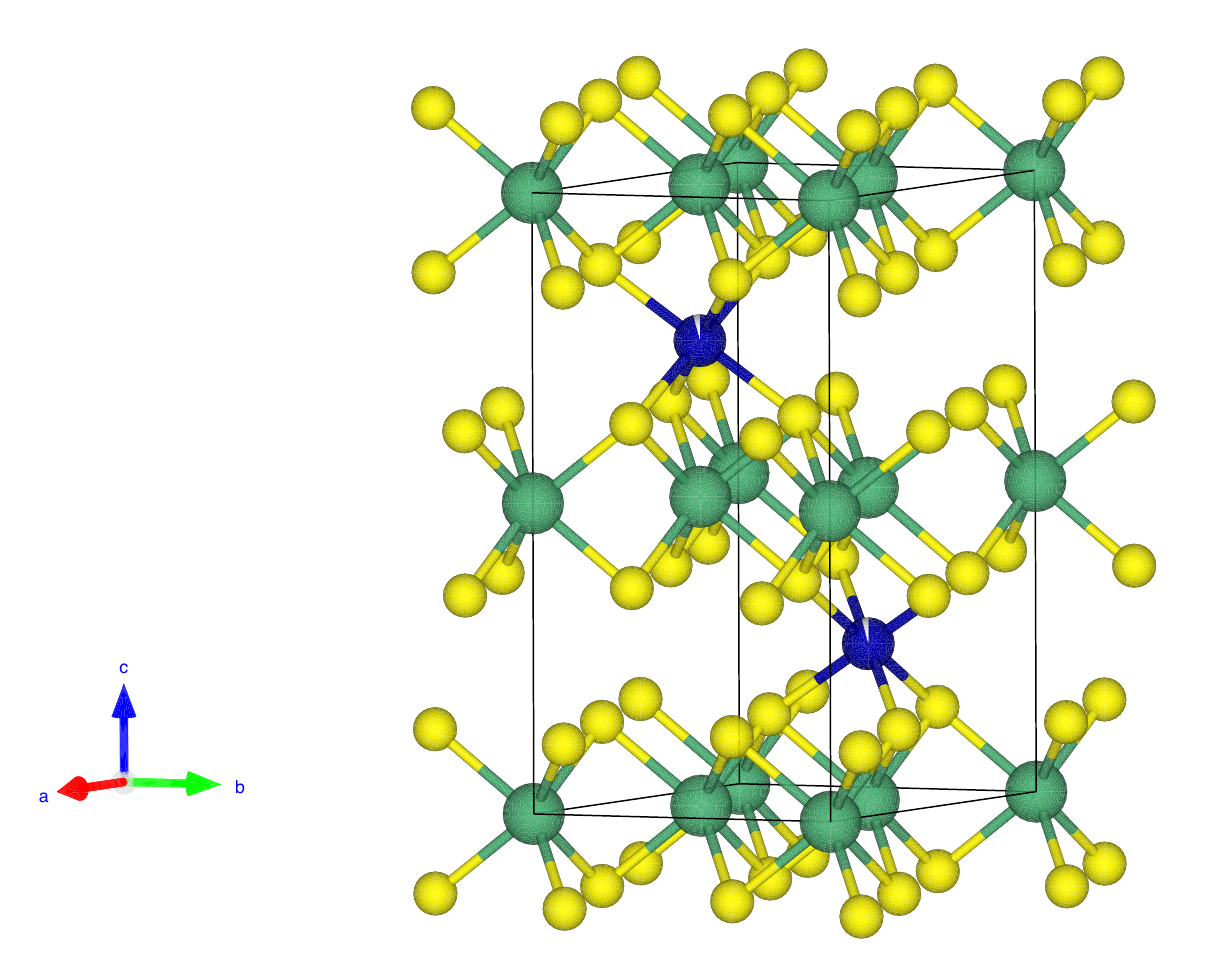}
    \caption{Unit cell of CoNb$_3$S$_6$ with Co in blue, Nb in green, and S in yellow, with the directions of lattice vectors $a$ (red), $b$ (green), and $c$ (blue) indicated.}
    \label{fig:CoNb3S6}
\end{figure}

A family of triangular magnets are the intercalated (TM)Nb$_3$S$_6$ compounds, where TM is Ni, Co, Fe, or Mn. These materials are dichalcogenides\cite{anzenhofer} (TM)$_{x}$NbS$_2$ in which TMs are intercalated between prismatic layers of NbS$_2$ and are stable and ordered at $x=1/3$. The unit cell for CoNb$_3$S$_6$ is depicted in Fig.~\ref{fig:CoNb3S6}. The crystal structure and magnetic susceptibilities were first investigated by Anzenhofer {\em et al.}\cite{anzenhofer}, who also discussed their electronic structure. Parkin, Marseglia, and Brown\cite{parkin1983magnetic} used neutron diffraction on single-crystal samples of CoNb$_3$S$_6$ and CoTa$_3$S$_6$ to determine the AFM magnetic structure of these two compounds. They concluded that the magnetic structure is orthohexagonal with two Co atoms per magnetic unit cell, with a moment of 2.73~$\mu_B$ for Co, slightly lower than the spin-only moment of 3~$\mu_B$ for Co$^{2+}$. More recently, Ghimire {\em et al.}\cite{ghimire2018large} performed magnetic measurements and magnetotransport measurements on CoNb$_3$S$_6$. They found a small linear susceptibility for in-plane and out-of-plane magnetic fields (less than  0.1~$\mu_B$ per formula unit for a field of 6~T), with the out-of-plane susceptibility larger than the in-plane one, but with a pronounced hysteresis in the out-of-plane susceptibility at temperatures below 29~K. Magnetotransport measurements yielded a relatively large AHE below the N\'eel temperature $T_N$, which is $27.5$~K. They argued that a magnetic field-induced component out of plane was not large enough to give rise to the observed AHE. Based on electronic structure calculations, they suggested that CoN$_3$S$_6$ is a magnetic Weyl semimetal with a complex non-collinear magnetic structure. Later, Tenasini {\em et al.}\cite{tenasini2020giant} performed further magnetotransport experiments and found an AHE per Co-layer close to the quantized value of $e^2/h$, suggesting that the Co-layers form topologically nontrivial 2D bands. 

The precise magnetic structure of (TM)Nb$_3$S$_6$ remains elusive, but the works by Ghimire\cite{ghimire2018large} and Tenasini\cite{tenasini2020giant}, in particular, suggest that there is a connection between the magnetic structure and the electronic structure, giving rise to non-trivial topology and a large AHE. In the present work, we look for magnetic order as a possible source for a non-zero AHE. We propose magnetic ground states for a model of the (TM)Nb$_3$S$_6$ systems. We find that, depending on the ratio of coupling constants, the ground state can be either a non-collinear, non-coplanar AFM with non-zero chirality, or spiral states $1q$ and $2q$ defined, respectively, by a single wavevector ${\mathbf q}$ or by two wavevectors $\mathbf{q}$ and ${\mathbf q}_2$ in the first Brillouin Zone (BZ), and in particular $\mathbf{q}_2$ is in general incommensurate with the in-plane lattice constant. In addition, the spiral states have zero chirality: the $1q$ state is invariant under the combination of ${\mathcal T}$ and ${\mathcal R}$. 
This makes the contribution to the AHE from a real-space chirality and the Berry phase vanish, and the $1q$ magnetic state in (TM)Nb$_3$S$_6$ cannot yield a nonzero AHE\cite{takatsu2010,chen2014anomalousMn3Ir}. The $2q$ state is more complicated: it is in general non-coplanar and has a local chirality that does not vanish, but the average chirality over many plaquettes vanishes, which implies that the AHE will, too. This means that the AHE can be a discriminant of the magnetic ground states. The ground state we find for a range of interaction parameters is consistent with the ground state obtained from electronic structure calculations including spin-orbit interactions\cite{hyowon}. The paper is organized as follows. In Sec.~\ref{sec:methods} we introduce the magnetic Hamiltonian, and in \ref{sec:results} we discuss finite-temperature atomistic simulations, and introduce general variational ground states for the different candidate states. We present our results in  section~\ref{sec:results}, and  Sec.~\ref{sec:discussion} contains conclusions and summary.

\begin{figure}
    \centering
    \includegraphics[width=0.85\columnwidth]{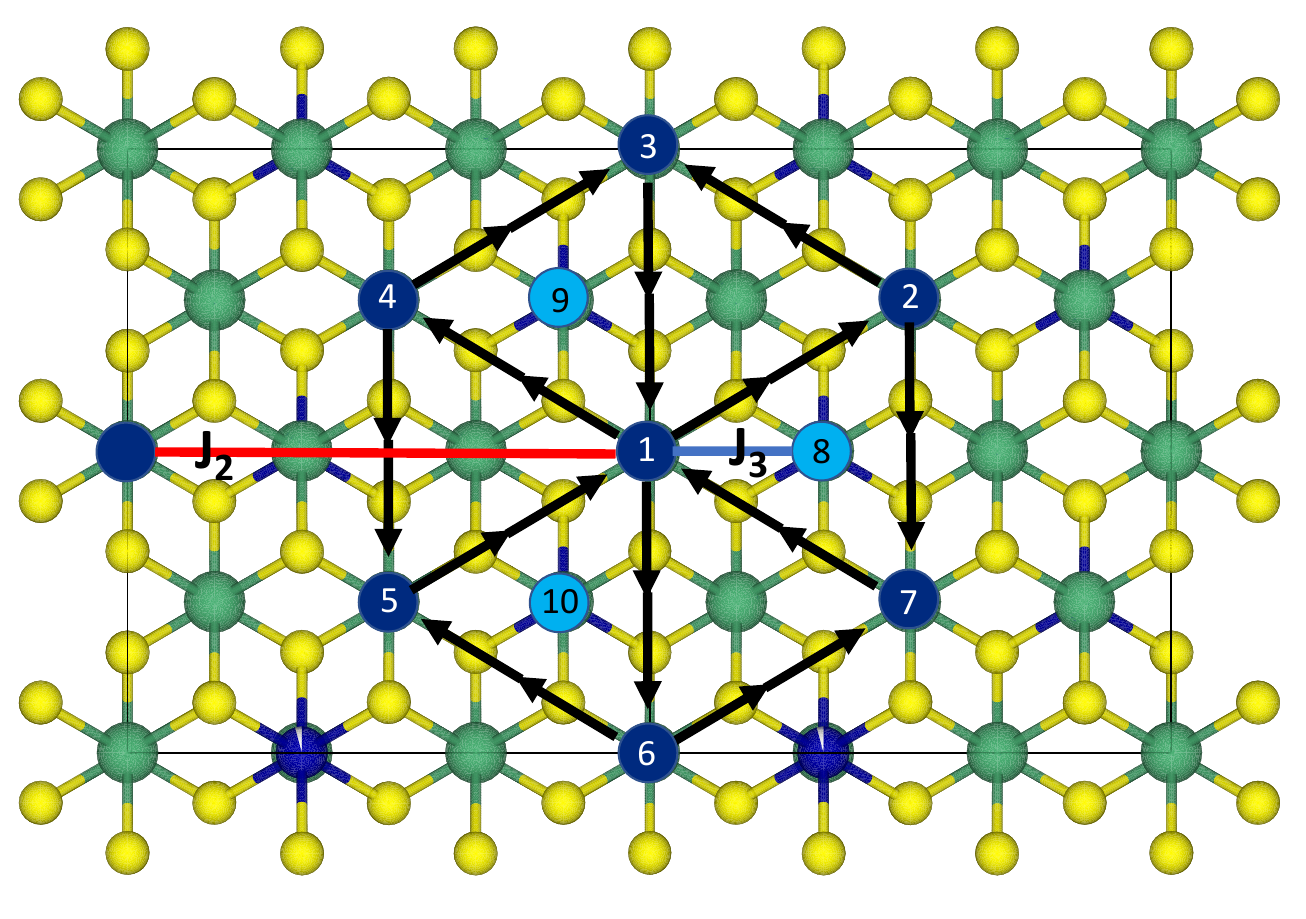}
    \caption{View of CoNb$_3$S$_6$ in the $ab$-plane. Co atoms in dark blue are on one lattice plane along the $c$ axis, and light blue ones on neighboring planes. In the plane, the Co atom labeled 1 interacts with the in-plane nearest neighbors 2 - 7, with in-plane next-nearest neighbors as indicated by the red line, and with  out-of-plane neighbors at the sites 8 to 10 as indicated by a blue line. The arrows on the bonds indicate the order of the cross product for the near-neighbor DMI.}
    \label{fig:CoNb3S6_bonds}
\end{figure}
\section{Methods}\label{sec:methods}
\subsection{Model Hamiltonian}\label{subsec:hamiltonian}

Experimental evidence makes clear that (TM)Nb$_3$S$_6$ undergo magnetic transitions from paramagnetic to ordered magnetic states\cite{anzenhofer,parkin1983magnetic,ghimire2018large,tenasini2020giant} at temperatures of about 30~K or higher. First-principle calculations\cite{hyowon} yield TM moments ranging from 1.4~$\mu_B$ (TM=Ni) to 4.9~$\mu_B$ (TM=Mn), and experimental measurements\cite{parkin1983magnetic} also indicate a large Co moment of 2.73~$\mu_B$.  These are temperature ranges and magnetic moments for which classical spin models are usually applied successfully. We are furthermore not aware of any evidence that quantum spin fluctuations play an important role in the magnetic structure or transport measurements. We will therefore use classical spin models to describe these systems. We assume that the magnetization can be described by local moments on the TM atoms and start with a minimal classical Heisenberg model with near-neighbor in-plane AFM coupling $J$ and biquadratic coupling $B$, near-neighbor out-of-plane (OOP) coupling $J_3$ (see Fig.~\ref{fig:CoNb3S6_bonds}). The moments are located on a triangular lattice in the crystallographic $ab$-plane, which we will take to be the $xy$-plane, with lattice constant $a$, and we take the $z$ axis to be along the crystallographic $c$ axis, so the sites of the Co atoms are given by
\begin{equation}
    {\mathbf r}_i=m_i \left( \frac{\sqrt{3}}{2}a\hat x+\frac{a}{2}\hat y\right)+n_i a\hat y+\ell c\hat z+{\rm Mod}(\ell,2)\frac{\sqrt{3}}{4}a\hat x,
\end{equation}
where $m_i$, $n_i$, and $\ell$ are integers.
Because inversion symmetry is broken, a Dzyaloshinksii-Moriya interaction (DMI) is allowed, with the general form
\begin{equation}
    H_{\rm DMI}=\sum_{<i,j>}{\mathbf d}_{ij}\cdot\left[{\mathbf S}({\mathbf r}_i)\times {\mathbf S}({\mathbf r}_j)\right],
    \label{eqn:general_DMI}
\end{equation}
where the sum $<i,j>$ is over in-plane nearest neighbors on sites ${\mathbf r}_i$ and ${\mathbf r}_j$. 
Based on symmetry, the DMI vector must be directed along the crystallographic $c$ axis. 
There are then two possible ways to arrange the DMI vectors, along the $+z$ axis or along the $-z$ axis. 
We do not know if the DMI vectors point up or down but for the purposes of our work here, which one is lower in energy is immaterial, and we will take the DMI vectors to point up. We will also assume that there is a single-site anisotropy with the $ab$-plane an easy plane, consistent with the experimentally observed larger out-of-plane susceptibility than in-plane one\cite{ghimire2018large}. The 3D classical Hamiltonian is then
\begin{equation}
    {\mathcal H}_{3D}=H_{\rm exchange}+H_{\rm nn\,ex}+H_{\rm OOP}+H_{\rm DMI}+H_{\rm biq}+H_{\rm ani}+H_{\rm Z}.
    \label{eqn:H_3D}
\end{equation}
The nearest-neighbor in-plane exchange interaction is
\begin{equation}
H_{\rm exchange}=\frac{J}{2}\sum_{<i,j>}\mathbf{S}(\mathbf{r}_i)\cdot\mathbf{S}(\mathbf{r}_j),
\end{equation}
where we will take $J$ to be unity and to be the energy scale.  
We include an in-plane next-nearest neighbor exchange
\begin{equation}
    H_{\rm nn\,ex}=\frac{J_2}{2}\sum_{<<i,j>>}\mathbf{S}(\mathbf{r}_i)\cdot\mathbf{S}(\mathbf{r}_j),
\end{equation}
where the notation $<<i,j>>$ means that $i$ and $j$ are in-plane next-nearest neighbors.  
The near-neighbor out-of-plane exchange is
\begin{equation}
H_{\rm OOP}=\sum_{<i,j>,OOP}\frac{J_3}{2}\mathbf{S}(\mathbf{r}_i)\cdot\mathbf{S}(\mathbf{r}_j),
\end{equation}
with the sum over out-of-plane near-neighbor sites $i$ and $j$. We will assume that $J_3<J$. This is not unreasonable as the OOP bond length is larger than the in-plane one by about 1~{\AA}. This is in any case not important as $J_3$ just sets a scale for the inter-plane order which, as we show below (Eqs.~(\ref{eqn:OOP_1}) and (\ref{eqn:OOP_2})), is commensurate with the lattice spacing $c$.

The DMI is
\begin{equation}
H_{\rm DMI}=\sum_{<i,j>}D\mathbf{d}\cdot [\mathbf{S}(\mathbf {r}_i)\times\mathbf{S}(\mathbf {r}_j)],
\end{equation}
where $D$ is the coupling strength and $\mathbf{d}=\hat z$ is the DMI vector. 
The biquadratic exchange and uniaxial anisotropy are, respectively,
\begin{equation}
    H_{\rm biq}=\frac{B}{2}\sum_{<i,j>}\left[\mathbf{S}(\mathbf{r}_i)\cdot\mathbf{S}(\mathbf{r}_j)\right]^2,
\end{equation}
and 
\begin{equation}
    H_{\rm ani}=K\sum_i (S_{z,i})^2,
\end{equation}
with $B$ and $K$ the respective coupling strengths and $K>0$ for the $ab$-plane an easy plane. 
In addition, in the presence of an external field ${\mathbf H}_{\rm ext}$ there is a Zeeman energy
\begin{equation}
    H_{\rm Z}=-\sum_i\mathbf{H}_{\rm ext}\cdot\mathbf{S}(\mathbf{r}_i).
\end{equation}
Figure~\ref{fig:CoNb3S6_bonds} depicts a view of CoNb$_3$S$_6$ in the $ab$-plane. Co atoms in one plane along the $c$ axis are indicated in dark blue. The Co at site 1 interacts with its nearest neighbors on sites 2 to 7 via the Heisenberg interaction and the DMI; the order for the cross product in the DMI in elementary triangular plaquettes is indicated with the black arrows. The Co at site 1 also interacts with its in-plane next-nearest neighbor through a coupling $J_2$, as indicated by the read line. The sites colored light blue and labeled 8 to 10 are Co atoms in the plane above or below site 1, and the Co atom at site 1 interacts with these six sites through a coupling $J_3$ indicated with a blue line. 

It is not impossible that there are longer-range interactions in these compounds. For example, because (TM)Nb$_3$S$_6$ are metallic, there may be long-range Ruderman-Kittel-Kasuya-Yosida (RKKY) interactions mediated by electrons at the Fermi surface, and such interactions could lead to longer-range order such as spiral structures along the $c$ axis. However, because of the intercalated nature of these compounds with a large distance between consecutive TM planes, the out-of-plane resistivity (along the $c$ axis) is more than an order of magnitude larger than the in-plane resistivity\cite{tenasini2020giant}. This makes RKKY interactions along the $c$ axis unlikely to be large enough to have a significant effect. One may of course include more in-plane couplings. However, our model already includes four in-plane couplings that extend up to (10~{\AA}) through the next-nearest neighbor interactions. We are also interested in a minimal model that can explain the magnetic structures and the appearance of a large AHE in CoNb$_3$S$_6$, and we believe our model can. We will therefore not try to extend it to include more interactions (which would also necessarily make analyses more complicated).

Given the Hamiltonian $\mathcal{H}_{3D}$ in Eq.~(\ref{eqn:H_3D}), there are a few properties of the magnetic order one may expect. Because the system is a triangular antiferromagnet with ABAB stacking, the system should have a simple commensurate order along the $c$-axis\cite{reimers1992}. A simple argument illustrates this: Given the structure of the Hamiltonian Eq.~(\ref{eqn:H_3D}) with decoupled in-plane and OOP couplings, one can assume that the spin configuration in an ordered state is separable into in-plane and OOP components, and the latter can be Fourier transformed:
\begin{equation}
{\mathbf S}({\mathbf r}_i)=\frac{1}{N_z}\sum_{q_z}{\mathbf S}(x_i,y_i,q_z)e^{iq_z z_i},
   \label{eqn:OOP_1}
\end{equation}
where $N_z$ is the number of TM planes. 
This immediately leads to an effective OOP coupling by summing over the six OOP near-neighbors that couple to the spin at site ${\mathbf r}_i$
\begin{equation}
    H_{OOP}=\frac{J_3}{N_z}\sum_{q_z,<i,j>, OOP}\cos(q_zc)
   {\mathbf S}(x_i,y_i,q_z)\cdot {\mathbf S}^*(x_j,y_j,q_z),
   \label{eqn:OOP_2}
    \end{equation}
which is minimized for $q_z=0$ ($J_3<0$, ferromagnetic OOP coupling) or $q_z=\pi/c$ ($J_3>0$, AFM OOP coupling), as the in-plane couplings $J$, $J_2$, and $B$ are all antiferromagnetic, and, as we shall argue later at the end of Sec.~\ref{sec:results} A, $D$ must be small. This means that we can expect the order along the $c$ axis to be trivial, irrespective of the sign of $J_3$. Furthermore, given the nature of the DMI, we expect that increasing the DMI will tend to make the in-plane spin order coplanar, at least for spin spiral state, in order to minimize the DMI energy. Finally, the next-nearest neighbor interaction can lead to an instability of the in-plane static susceptibility at the M points in the BZ\cite{martin2008itinerant,solenov2012}, which can potentially lead to the emergence of a $3q$ state. The $3q$ state is a non-coplanar AFM with a non-zero chirality; such a state can give rise to a non-zero AHE because the non-zero chirality corresponds to a Berry phase\cite{martin2008itinerant}.

\begin{figure}
\centering
    \includegraphics[width=0.95\columnwidth]{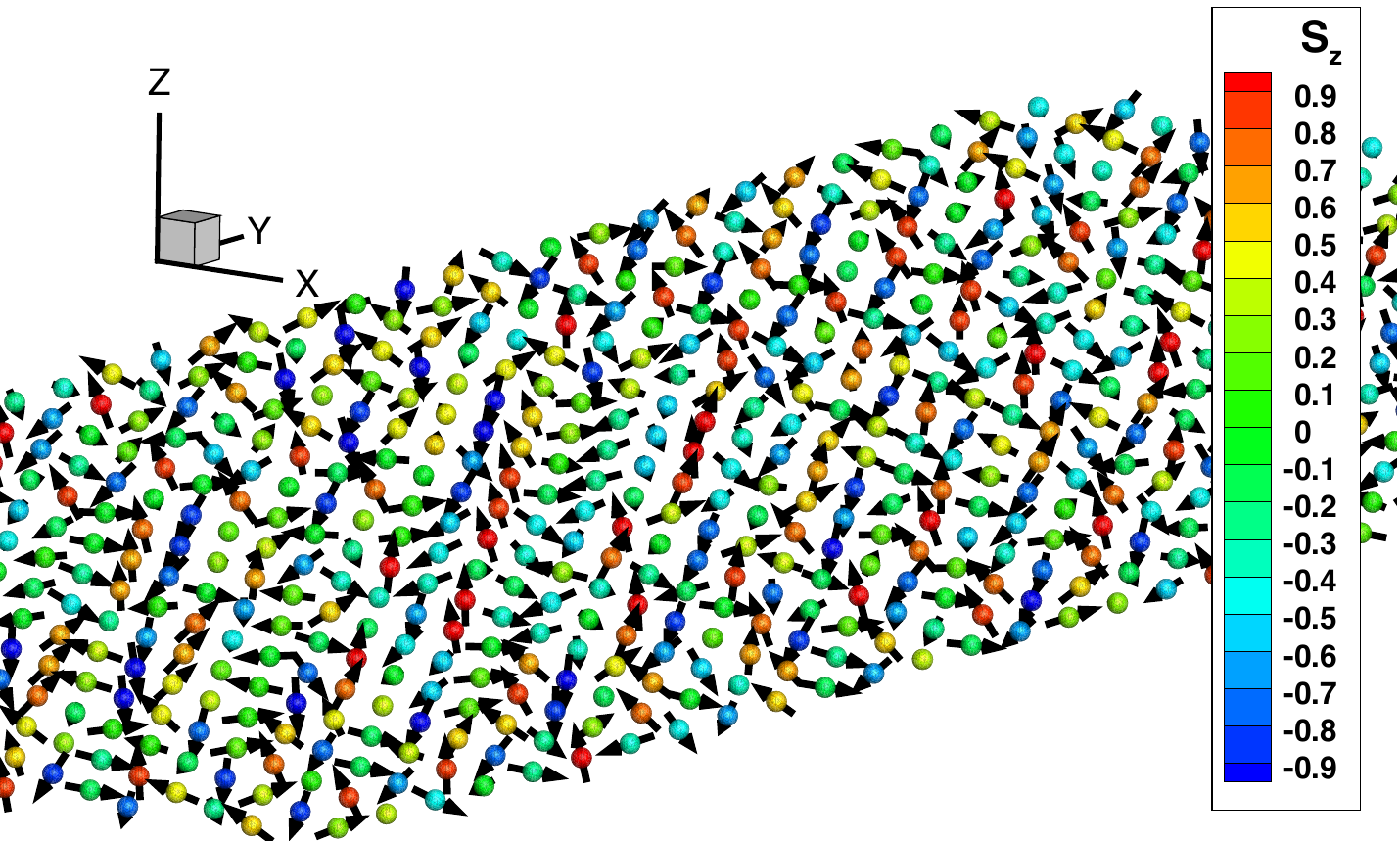}
    \caption{Snapshot of the spin configuration in a TM plane for $B=0.4$, $K=0.1$, $D=0$, $J_2=0.08$, $J_3=0.2$, and $k_BT\approx0.01$. The arrows show 3D the spin orientation, and the color coding denotes the $z$-component of the spins. The snapshot shows there is short-range local order in small domains.}
    \label{fig:B04J2008}
\end{figure}

\begin{figure*}
\centering
    \includegraphics[width=3.5 truein]{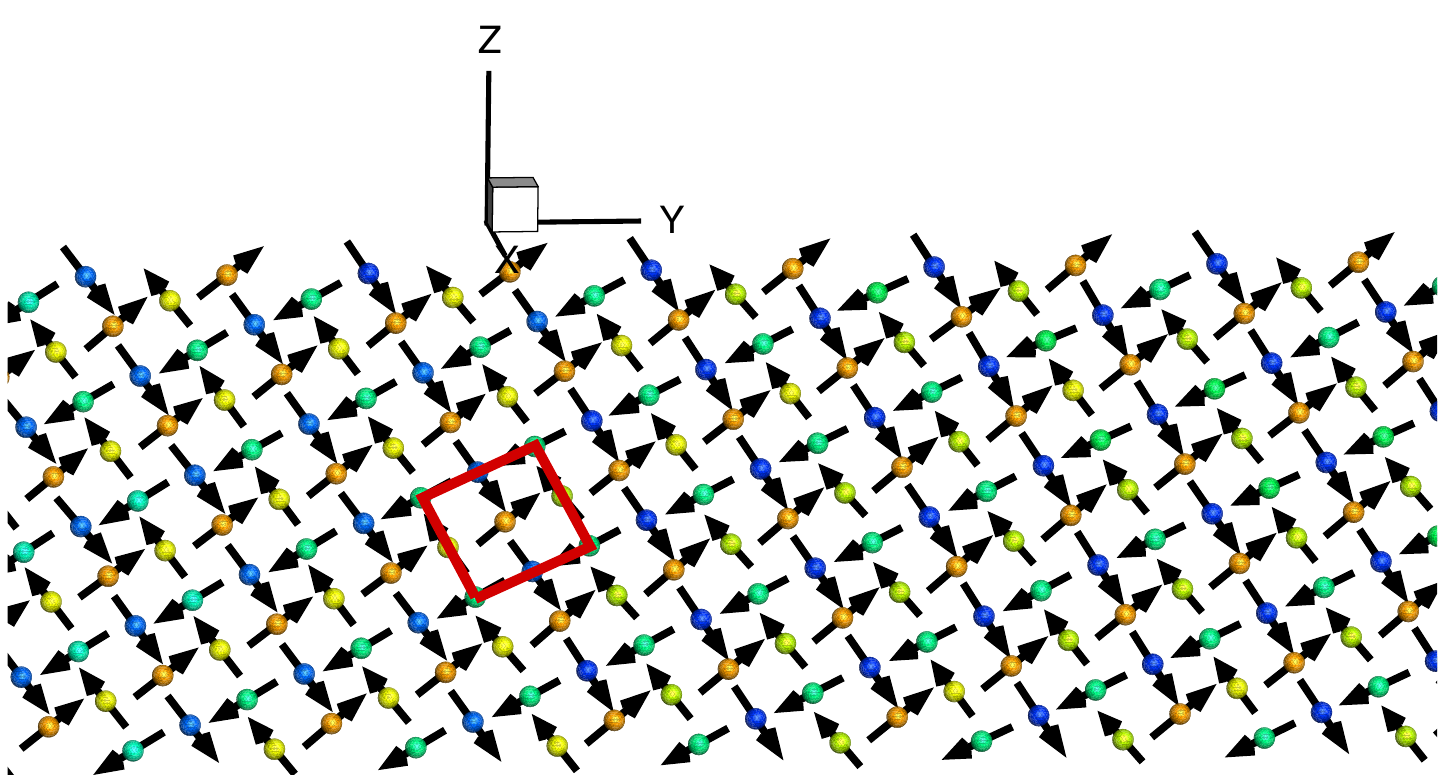}
    \includegraphics[width=3.5 truein]{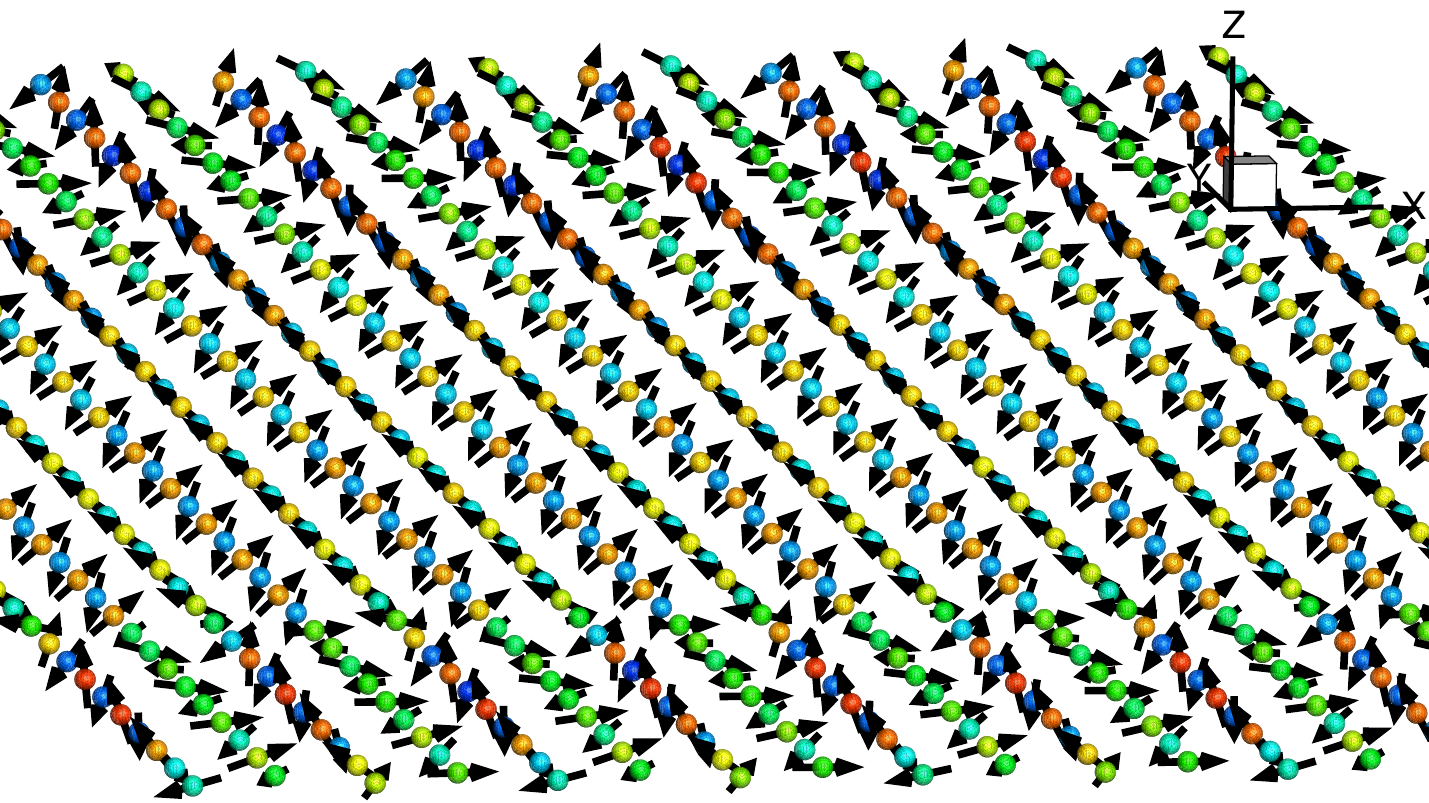}
    \caption{Snapshots of the spin configuration in a TM plane for $B=0.4$, $K=0.1$, $D=0$, $J_3=0.2$, $J_2=0.3$, (left panel), and $J_2=0.5$ (right panel), and $k_BT\approx0.01$. The arrows show 3D the spin orientation, and the color coding denotes the $z$-component of the spins with the same color scale as in Fig.~\ref{fig:B04J2008}. For $J_2=0.3$, the spins depicted are almost in a single-domain $3q$ state (there is a domain wall towards the right end of the figure). In the right panel ($J_2=0.5$) several domains are visible. The largest domain in the center of the figure is not a $3q$ state as a clear twist of the spins is visible along the $y$-axis; this is probably a $2q$ state in which $\mathbf{q}_2$ is incommensurate with the lattice.}
    \label{fig:B04J2_2}
\end{figure*}

In order to establish some basic features of the  the low-temperature magnetic structure of ${\mathcal H}_{3D}$ in Eq.~(\ref{eqn:H_3D}), in particular to confirm the trivial out-of-plane order, we performed Monte Carlo simulations and also simulations integrating the stochastic Landau-Lifshitz-Gilbert (s-LLG) equation\cite{garcia1998langevin} based on the Hamiltonian ${\mathcal H}_{3D}$ at fixed temperature using the Vampire software\cite{Vampire}. For the fixed-temperature s-LLG simulations, we used a timestep of 0.1~fs and a dimensionless damping $\alpha=0.1$, and thermally randomized the spins at a high temperature $k_BT\approx 1$ for 1~ns ($10^5$ time steps), and then quenched the system to a low temperature $k_BT\approx0.01$. We used an orthorombic supercell with dimensions 9.99047~nm$\times$23.0720~nm$\times$11.886~nm containing 16,000 atoms. As one might expect, short-range in-plane order emerged at low temperatures $T\sim J$ and long-range order at a temperature set by $J_3$. A main conclusion of these 3D simulation was that the order along the $c$ axis was always trivial, as the arguments above suggest, whether or not $J_3$ was ferromagnetic ($J_3<0$) or antiferromagnetic ($J_3>0$) with consecutive planes along $c$ having the same in-plane order shifted by an in-plane translation:  
the OOP coupling leads to a trivial order along the $c$ axis, without any effect on the in-plane order. Figure \ref{fig:B04J2008} shows a snapshot of the spin configuration in an $ab$-plane for $B=0.4$, $J_2=0.08$, $D=0$, $J_3=0.2$ and $k_BT\approx0.01$. There appears to be some local order but there are multiple domains in the imaged region. It should be noted that at this low temperature, only very small thermal noise can be discerned as a function of time. The difficulty in identifying the nature of the order is often the case for finite-size simulations when the order may be incommensurate with the lattice spacing. Figure~\ref{fig:B04J2_2} shows snapshots of the spin configurations for the same parameters $B$, $D$, and $J_3$, but now with $J_2=0.3$ (left panel) and $J_2=0.5$ (right panel). In these figures spin ordering is clearly discernible, even though the right panel contains a domain wall. A closer examination of the configuration for $J_2=0.3$ suggests that the order is a $3q$ order\cite{martin2008itinerant}.

\begin{figure}
\centering
    \includegraphics[width=0.9 \columnwidth]{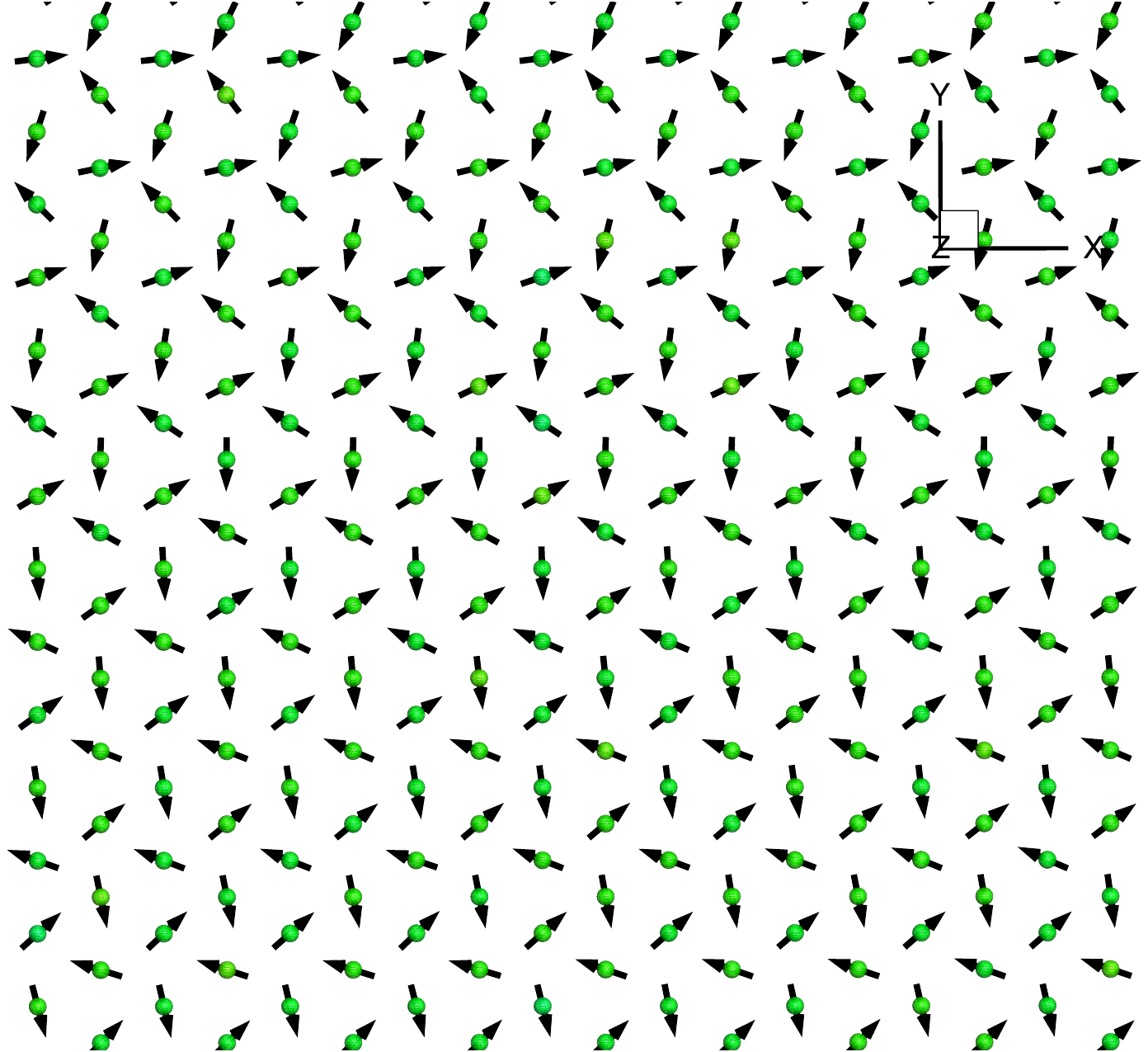}
    \caption{Snapshot of the spin configuration in a TM plane for $B=0.3$, $K=0.1$, $D=0$, $J_2=0.05$, $J_3=0.2$, and $k_BT\approx0.01$. The arrows show 3D the spin orientation, and the color coding denotes the $z$-component, which is here zero, of the spins with the same color scale as in Fig.~\ref{fig:B04J2008}. This state can readily be identified as a planar N\'eel state.}
    \label{fig:B03J2005}
\end{figure}

\begin{figure*}
\centering
    \includegraphics[width=3.5 truein]{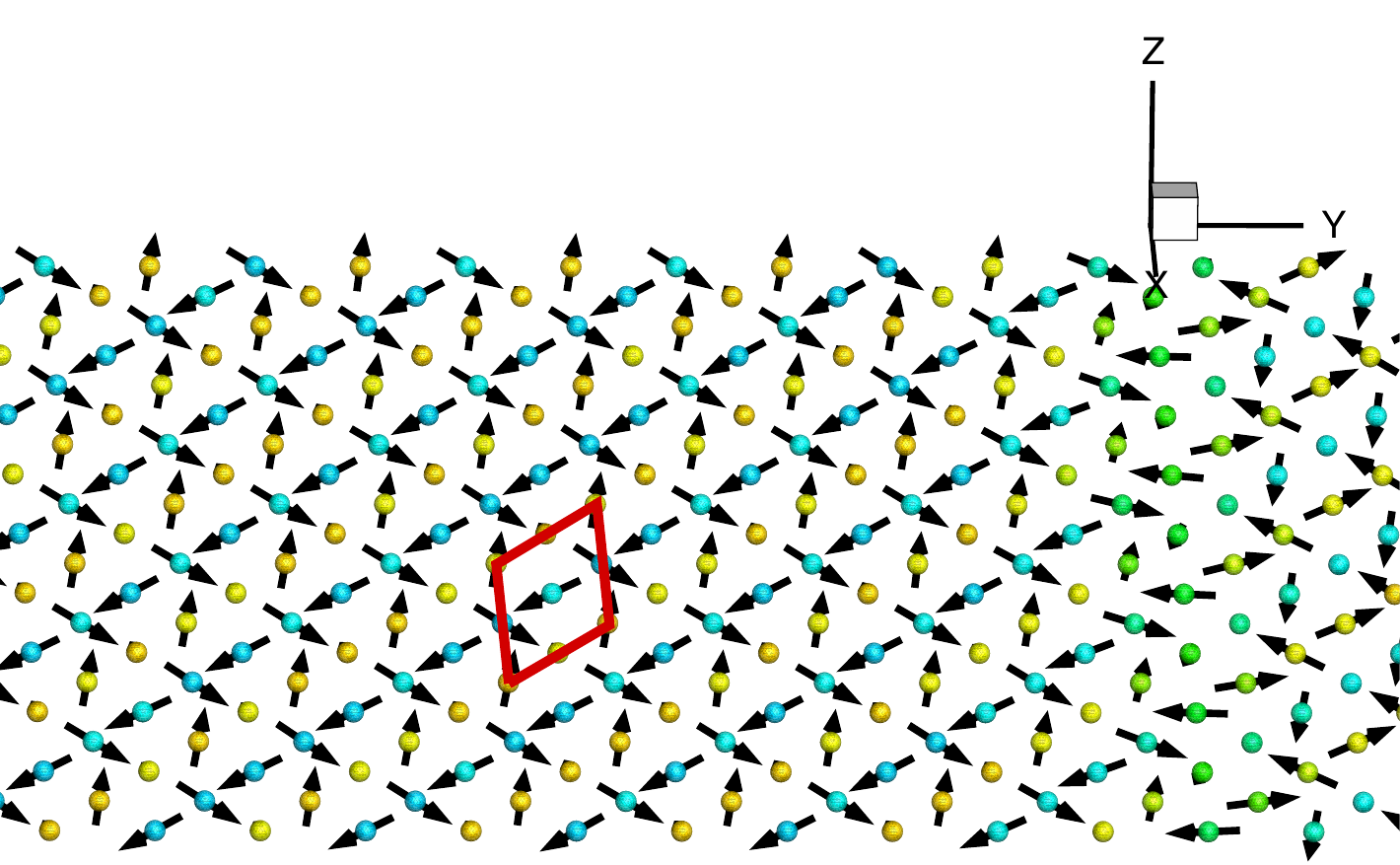}
    \includegraphics[width=3.5 truein]{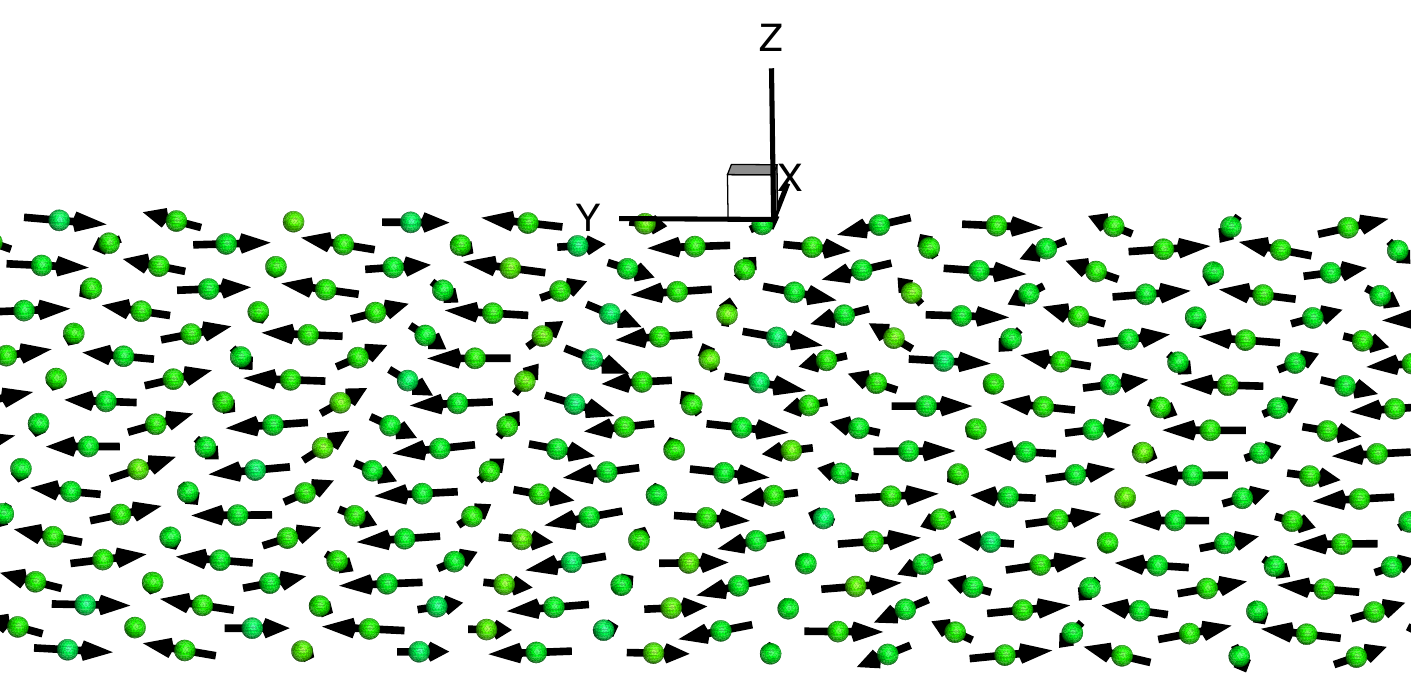}
    \caption{Snapshots of the spin configuration in a TM plane for $B=0.3$, $K=0.1$, $D=0$, $J_3=0.2$, $J_2=0.25$, (left panel), and $J_2=0.35$ (right panel), and $k_BT\approx0.01$. The arrows show 3D the spin orientation, and the color coding denotes the $z$-component of the spins with the same color scale as in Fig.~\ref{fig:B04J2008}. For $J_2=0.25$ (left panel) the state can be identified as a $3q$ state, and a magnetic unit cell is indicated with red lines. For $J_2=0.35$ (right panel) the order is more complicated and a gentle twist along the $x$ axis can be observed.}
    \label{fig:B03J2_2}
\end{figure*}

Figure~\ref{fig:B03J2005} shows a snapshot of the order in the $ab$ plane for $B=0.3$, $K=0.1$, $D=0$, $J_2=0.05$, $J_3=0.2$, and $k_BT\approx0.01$. In this figure, the order is clearly the classic N\'eel order, which can be described as a $1q$ order with the wavevector $q$ at a point K in the BZ. Figure~\ref{fig:B03J2_2} similarly shows snapshots for $J_2=0.25$ (left panel) and $J_2=0.35$ (right panel). At $J_2=0.25$, the system again exhibits the $3q$ state, while at $J_2=0.35$, the order has a short wavelength along the $y$ direction and a much longer wavelength is discernible along the $x$ direction, visible as a gentle twist of the spins.
\begin{figure}
\centering
    \includegraphics[width=0.40 \columnwidth]{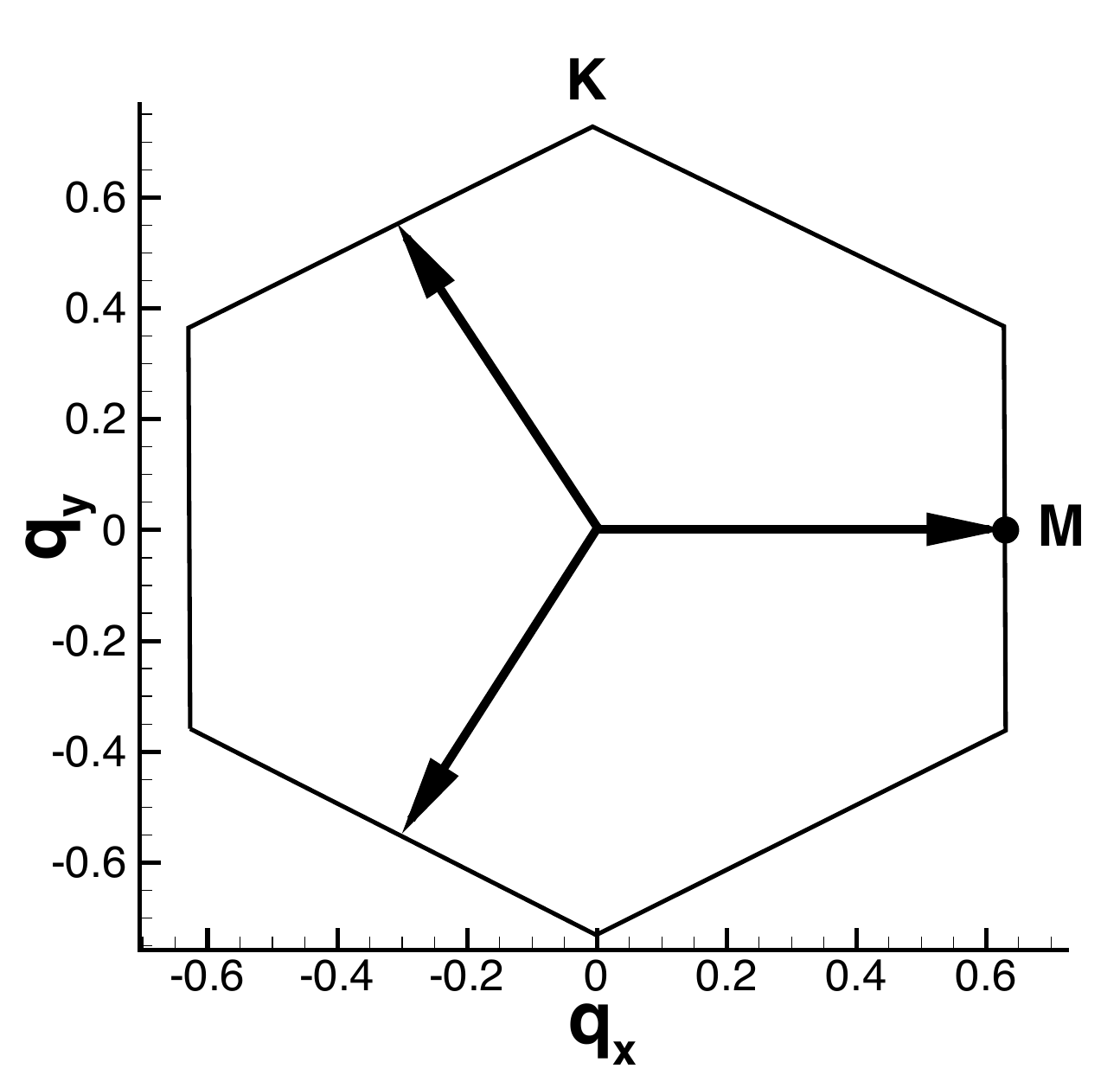}
    \includegraphics[width=0.45 \columnwidth]{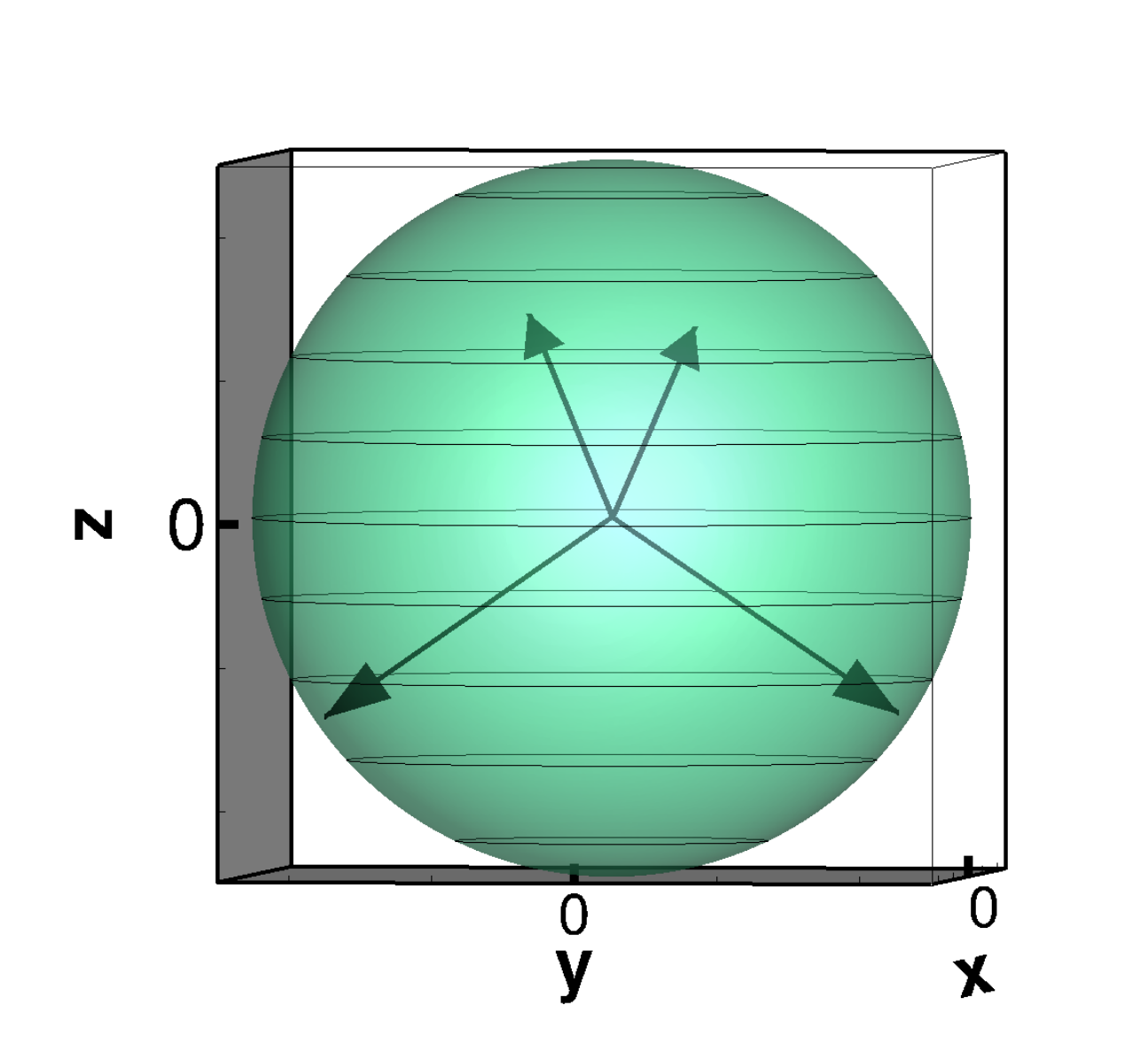}
    \caption{The left panel shows a cartoon of the three wavevectors of the $3q$ state. The three vectors extend from the zone center to three M points on the BZ boundary, $120^\circ$ degrees apart. The right panel shows a real-space depiction of the four spins in a magnetic unit cell. The four spins point towards four different corners in the spin-space unit cube, e.g. $(-1,1,1)$, $(1,-1,1)$, $(-1,-1,-1)$, and $(1,1,-1)$, such that the sum of the spins is zero.} 
    \label{fig:3Q_cartoons}
\end{figure}

The finite-T atomistic simulations thus confirmed the trivial OOP order, which allowed us to reduce the model to a two-dimensional (2D) in-plane model.
The 2D model is given by
\begin{equation}
    {\mathcal H}=H_{\rm exchange}+H_{\rm nn\,ex}+H_{\rm DMI}+H_{\rm biq}+H_{\rm ani}+H_{\rm Z}.
\end{equation} with lattice vectors which we write as
\begin{eqnarray}
\mathbf{b}_1 & = &\frac{\sqrt{3}}{2}a\hat x+\frac{a}{2}\hat y+0\hat z\nonumber\\
\mathbf{b}_2 & = &a\hat y\nonumber\\
\mathbf{b}_3 & = & c\hat z,
\end{eqnarray}
where $\mathbf{b}_3$ is irrelevant and $a=5.768$~{\AA}.
The reciprocal lattice vectors are then
\begin{eqnarray}
\mathbf{\overline{b}_1} & = & \frac{\mathbf{b}_2\times\mathbf{b}_3}{\mathbf{b}_1\cdot\left(\mathbf{b}_2\times\mathbf{b}_3\right)}= \frac{4\pi}{\sqrt{3}a}\hat x\nonumber\\
\mathbf{\overline{b}_2} & = & \frac{\mathbf{b}_3\times\mathbf{b}_1}{\mathbf{b}_1\cdot\left(\mathbf{b}_2\times\mathbf{b}_3\right)}=\frac{4\pi}{\sqrt{3}a}\left[\frac{\sqrt{3}}{2}\hat y-\frac{1}{2}\hat x\right]\nonumber\\
\mathbf{\overline{b}_3} & = & \frac{\mathbf{b}_1\times\mathbf{b}_2}{\mathbf{b}_1\cdot\left(\mathbf{b}_2\times\mathbf{b}_3\right)}=\frac{2\pi}{c}\hat z.\nonumber\\
\end{eqnarray}
We will also use the vectors
\begin{eqnarray}
\mathbf{\tau}_1 & = &\frac{\sqrt{3}}{2}a\hat x+\frac{a}{2}\hat y+0\hat z\nonumber\\
\mathbf{\tau}_2 & = &-\frac{\sqrt{3}}{2}a\hat x+\frac{a}{2}\hat y+0\hat z\nonumber\\
\mathbf{\tau}_3 & = &-a\hat y
\end{eqnarray}
that connect nearest-neighbor sites in an elementary triangular plaquette, with directions given by the DMI bonds in Fig.~\ref{fig:CoNb3S6_bonds}. 

We will seek ground states among different classes of ordered state by constructing different {\em Ans\"atze} with variational parameters and minimizing the total energy with respect to those parameters. The variational states cover very general states with $1q$ and $2q$ orders, and also include generalizations of the non-coplanar $3q$ state. While one can in general look for states the order of which are characterized by multiple wavevectors using systematic Fourier expansions\cite{leonov2015,liu2016}, such expansions can typically be terminated after two components as the weights of higher-order components decay exponentially and do not give rise to any physically meaningful effects\cite{liu2016}. We will therefore not construct higher-ordered states than the $1q$, $2q$, and the $3q$ non-coplanar states as we believe these suffice to characterize the phase diagram of our model.  

A $1q$ ordered state can be described by the {\em Ansatz}
\begin{widetext}
\begin{equation}
  \mathbf{S}_{1q}({\mathbf r}_i)  =  
    \left(A\cos(\mathbf{q}\cdot\mathbf{r}_i+\varphi),
    \sin(\mathbf{q}\cdot\mathbf{r}_i+\varphi),
    \sqrt{1-A^2}\cos(\mathbf{q}\cdot\mathbf{r}_i+\varphi)\right),
    \label{eqn:1q_A1}
\end{equation}
\end{widetext}
where $-1\leq A\leq1$, $\varphi$ is an arbitrary phase with $0\leq\varphi\leq2\pi$, and $\mathbf q$ is any wavevector in the 2D BZ. This {\em Ansatz} obviously preserves normalization of the spin at each site. 
Note that there are two other possibilities, 
\begin{widetext}
\begin{equation}
    \mathbf{S}(\mathbf{r}_i)=\left(A\cos(\mathbf{q}\cdot\mathbf{r}_i+\varphi),\sqrt{1-A^2}\cos(\mathbf{q}\cdot\mathbf{r}_i+\varphi),\sin(\mathbf{q}\cdot\mathbf{r}_i+\varphi)\right),
    \label{eqn:1q_A2}
\end{equation}
\end{widetext}
and
\begin{widetext}
\begin{equation}
    \mathbf{S}(\mathbf{r}_i)=\left(\sin(\mathbf{q}\cdot\mathbf{r}_i+\varphi),A\cos(\mathbf{q}\cdot\mathbf{r}_i+\varphi),\sqrt{1-A^2}\cos(\mathbf{q}\cdot\mathbf{r}_i+\varphi)\right).
    \label{eqn:1q_A3}
\end{equation}
\end{widetext}
In order to further generalize the variational $1q$ spin states, we also perform a global $SO(3)$ rotation $\mathcal {R}(\theta_r,{\mathbf w})$ of all spins, where $\mathcal{R}(\theta_r,\mathbf{w})$ rotates the spin an angle $\theta_r$ about the unit vector $\mathbf {w}$: ${\mathbf S}({\mathbf r}_i)\to\mathcal{R}(\theta_r,\mathbf{w})\mathbf{S}(\mathbf{r}_i)$. This yields seven variational parameters, $A$, $q_x$, $q_y$, $\varphi$, $\theta_r$, $w_x$, and $w_y$. 
Finally, in the presence of an external magnetic field along the $z$ axis, we have to allow for a small $z$-component of the spins induced by the external field. We add this in the following way. We start with a given set of $A$, $q_x$, $q_y$, and $\varphi$, and then perform the $SO(3)$ rotation for a given $\theta_r$, $w_x$, and $w_y$ of all spins. We then add a small $z$-component $\delta z\ll1$ to all spins. This breaks the normalization of the spins, so a final step is to renormalize all spins by dividing each spin by its norm. In numerical optimizations of the $1q$ and $2q$ states with an applied field, we ensure that the field is small enough that the resulting component $\delta z$ is indeed smaller than 0.1.

Without an external magnetic field, we expect the three variational {\em Ans\"atze} Eqs.~(\ref{eqn:1q_A1}-\ref{eqn:1q_A3}) to be degenerate in energy, at least for $D=0$. This was indeed confirmed in the numerical minimization with respect to the variational parameters, and served as a convenient check on the numerical minimizations.

\begin{figure*}
    \centering
    \includegraphics[width=3.5 truein]{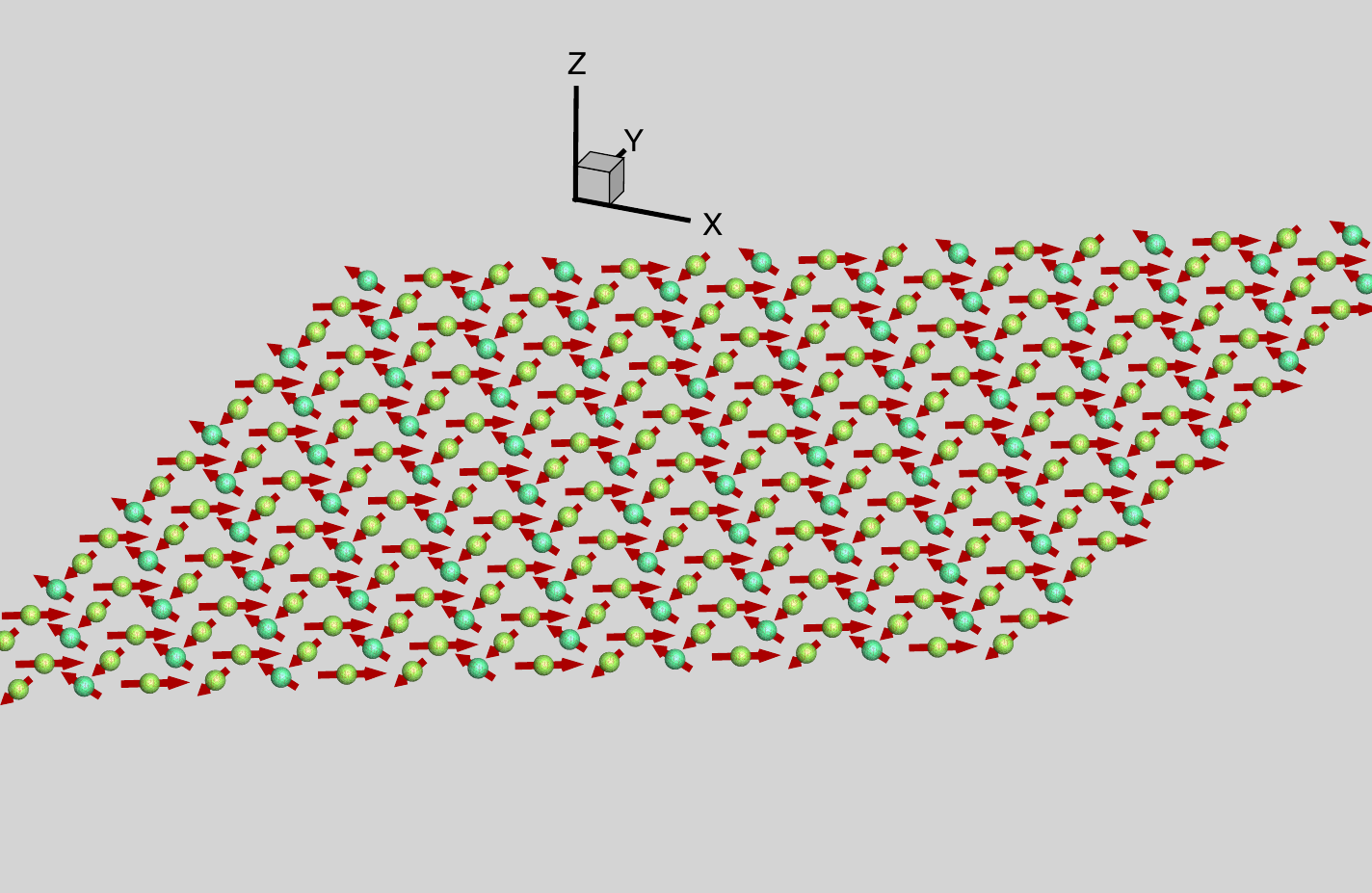}
        \includegraphics[width=3.5 truein]{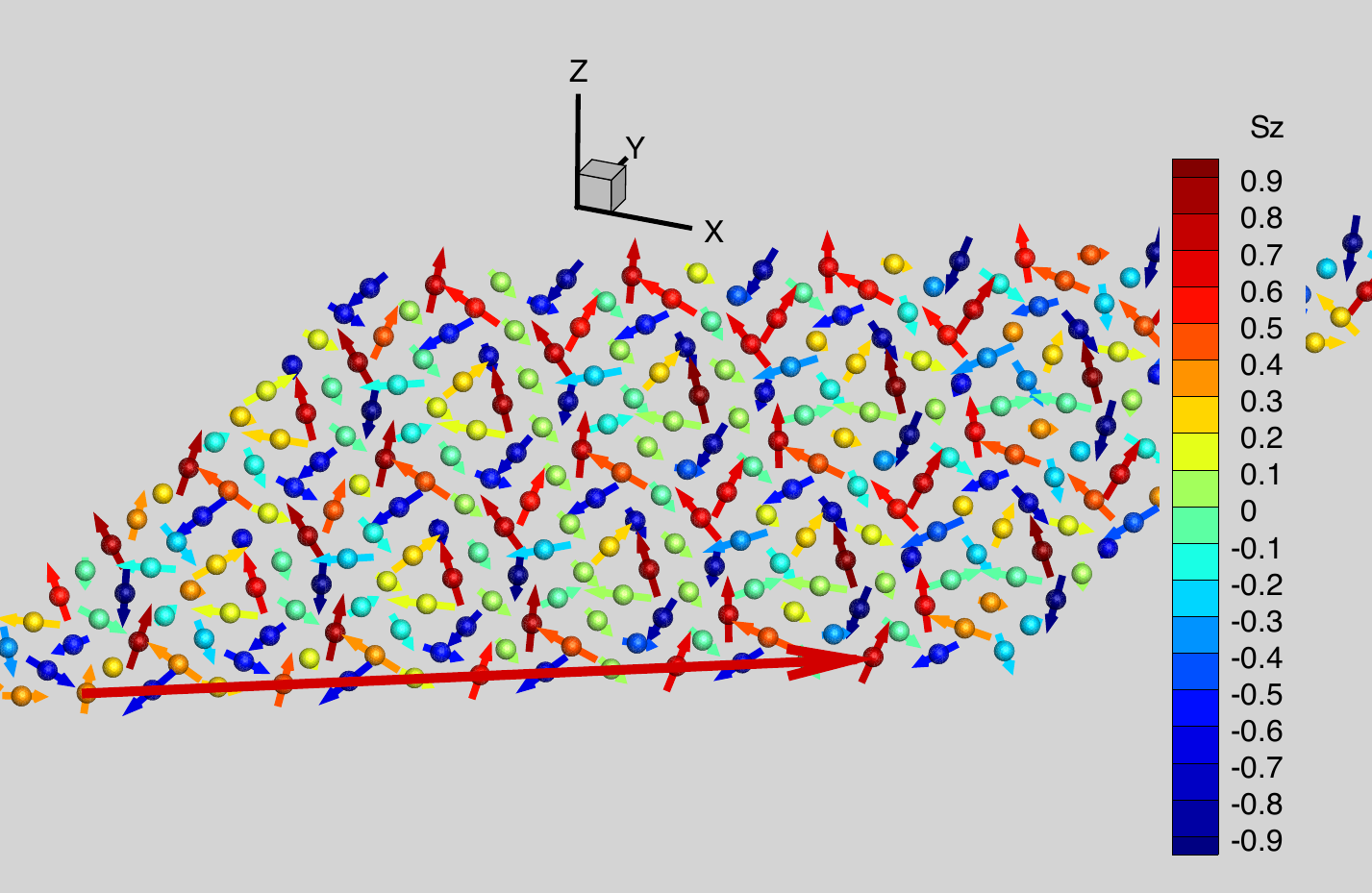}
    \caption{$1q$ (left panel) and $2q$ (right panel) variational solutions for $B=0.4$, $D=0$,  $J_2=0.04$ ($1q$), and $J_2=0.48$ ($2q$). The $1q$ state is a planar N\'eel state (the color coding of the arrows is in this case just for better visibility). In the right panel, a lattice direction along the $a$ axis is indicated. The spins have an almost commensurate periodicity of period three along this direction.}
    \label{fig:1q_2q_ex}
\end{figure*}

We construct variational $2q$ states by a simple generalization of the {\em Ans\"atze} Eqs.~(\ref{eqn:1q_A1}-\ref{eqn:1q_A3}) by replacing the constant amplitude $A$ and $\sqrt{1-A^2}$ for spin $\mathbf{S}(\mathbf{r}_i)$ by $\cos(\mathbf{q}_2\cdot\mathbf{r}_i)$ and $\sin(\mathbf{q}_2\cdot\mathbf{r}_i)$, respectively. Just as for the $1q$ variational states, we perform a global $SO(3)$ rotation for a given set of $q_x$, $q_y$, $q_{x2}$, $q_{y2}$, and $\varphi$. In the presence of an external magnetic field, we add a $z$-component $\delta z$ to all spins after the $SO(3)$ rotation, and then renormalize the spins. Note that these variational $2q$ states include the $1q$ ones as special cases. This provided another check on the numerical minimizations. 
Figure~\ref{fig:1q_2q_ex} shows examples of an optimized $1q$ spin state for $B=0.4$, $K=0.1$, $D=0$, and $J_2=0.04$, and an optimized $2q$ spin state for $B=0.4$, $K=0.1$, $D=0$, and $J_2=0.48$. The $1q$ spin state is planar. The $2q$ state is non-coplanar and has a rather complicated real-space texture, but certain features can be discerned. For example along the $a$ axis, as indicated in the figure, the spins have an almost commensurate period of three lattice spacings. It is not quite commensurate as the spins are twisted slightly away from each other at every third site.  

For the 2D model with near-neighbor, next-nearest-neighbor and biquadratic exchange and a small out-of-plane anisotropy (e.g., $K\approx0.05$ or $K\approx0.1$), with the $c$ axis a hard axis, there is a parameter range with $J_2$ smaller than unity where an in-plane so-called $3q$ state is the ground state\cite{martin2008itinerant}, in which the spin state is given by
\begin{equation}
    {\mathbf S}_{3q}({\mathbf r}_i)=
    ({\mathcal S}_1\cos({\mathbf q}_1\cdot{\mathbf r}_i),{\mathcal S}_2\cos({\mathbf q}_2\cdot{\mathbf r}_i),{\mathcal S}_3\cos({\mathbf q}_3\cdot{\mathbf r}_i)),
\end{equation}
where ${\mathcal S}_i$ are amplitudes with $\sum_{i=1}^3{\mathcal S}_i^2=1$. 
The vectors ${\mathbf q}_i$ extend from the $\Gamma$ point in the BZ to three $M$ points such that the ${\mathbf q}$-vectors are $120^\circ$ apart (see Fig.~\ref{fig:3Q_cartoons}). This yields a spin configuration with four inequivalent sites, so the magnetic unit cell has four sites (see Fig.~\ref{fig:B03J2_2}). In the absence of a DMI, the magnetization on the four sites are related by reflection or inversion, so there are only two degrees of freedom needed to specify the spin arrangement. These can be thought of as the magnitude of the spin projection on the $z$ axis, and a rotation about the $z$ axis. 
For $B=0$, the spin arrangement forms an a co-planar antiferromagnet with the amplitude of the $z$-component ${\mathcal S}_3=0$.  A small positive biquadratic coupling $B$ (in our case already for $B=0.025)$ can drive ${\mathcal S}_3$ non-zero, yielding a non-coplanar antiferromagnet. 
A non-zero DMI or an applied external field can potentially break the symmetry relations between the spin orientations on the four different sites; in particular, the magnitude of the $z$ component, ${\mathcal S}_3$, can on the sites with ${\mathcal S}_3<0$ be different from the sites with ${\mathcal S}_3>0.$ In order to allow for this possibility, we construct a bipartite model with positive and negative $z$-components of the spin and seek solutions of the form
\begin{widetext}
\begin{equation}
    {\mathbf S}({\mathbf r}_i)=
    (\cos(\varphi_j)\sin(\theta_j)\cos({\mathbf q}_1\cdot{\mathbf r}_i),\sin(\varphi_j)\sin(\theta_j)\cos({\mathbf q}_2\cdot{\mathbf r}_i),\cos(\theta_j)\cos({\mathbf q}_3\cdot{\mathbf r}_i))
\end{equation}
\end{widetext}
where $\theta_j$ and $\varphi_j$, $j=1,2$ are additional variational parameters, and $j$ enumerates the two sub-lattices with positive and negative $\mathcal{S}_3$. This {\em Ansatz} with four variational parameters then also allows for an out-of-plane net magnetization driven either by interactions or by an applied external field as the magnitude of the $z$-component of the spin can be different on one sublattice from the other.

\begin{figure}
    \centering
    \includegraphics[width=0.75 \columnwidth]{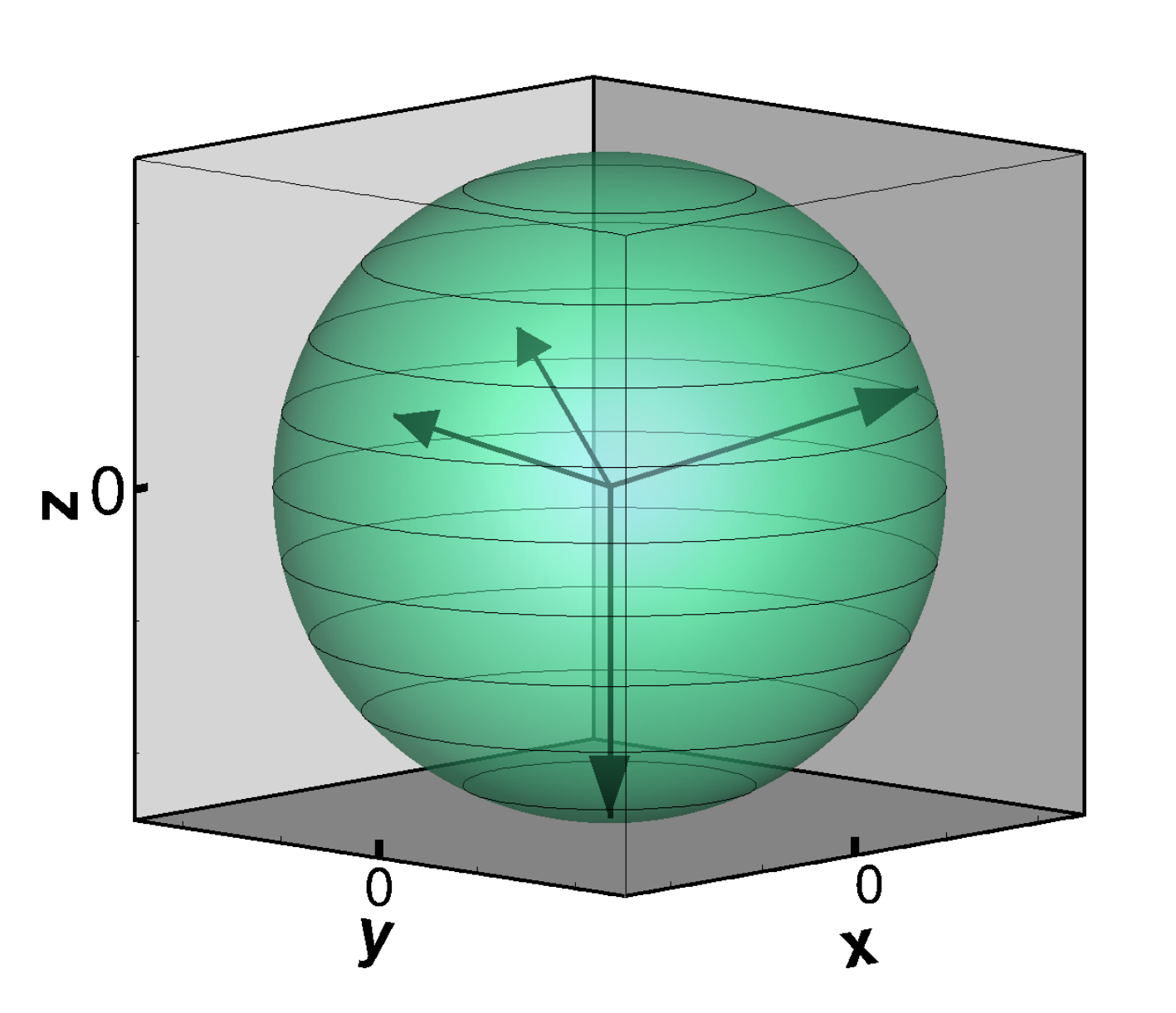}
    \caption{Depiction of another class of spin states with the four spins from one magnetic unit cell inserted in the unit cube in spin space. One spin  has positive or negative $z$-component (here shown with negative $z$ component), and the others have $z$-components of opposite sign and the spin components in the $xy$ plane $120^\circ$ apart.}
    \label{fig:mercedes}
\end{figure}

In the absence of DMI and for $K=0$, the $3q$ state is degenerate under arbitrary $SO(3)$ rotations of all spins. In the presence of DMI and anisotropy, this is generally no longer the case. 
However, with the $c$ axis a hard axis and with the DMI vector also along the $c$ axis, the Hamiltonian is invariant under arbitrary global spin rotations about the $c$ axis. 
There is then another readily identified spin  state with net zero magnetization compatible with the lattice symmetry. This state also has four spins per unit cell, with one spin, $\mathbf{S}_0$, along the $z$ axis or perpendicular to the $z$-axis, and the other three with components equal to $1/3$ in magnitude with opposite sign to the $z$- or in-plane component of $\mathbf{S}_0$, and with the components perpendicular to $\mathbf{S}_0$ $120^\circ$ apart (see Fig.~\ref{fig:mercedes} for a depiction with $\mathbf{S}_0$ along the $-z$ axis.) For the parameter range examined here, this state has higher energy than the  $3q$ state, and we will ignore it from now on.

For a given set of input parameters $J_2, D, B, K$ and external field $H_z\hat z$, we then minimize the total energy per spin with respect to the parameters $\theta_j$ and  $\varphi_j$ for the $3q$ state, and with respect to $A, q_x, q_y, \varphi, \theta_r, w_x, w_y$, and $\delta z$ for the $1q$ and $2q$ spin spiral states. Because the interaction energy between nearest neighbor spins only depends on their relative orientation, we can without loss of generality put one spin at the origin and calculate the interaction energy of this spin. For the $3q$ state, it suffices to calculate the total energy  (interaction, anisotropy, and Zeeman) of the four inequivalent spins in the magnetic unit cell. For the $1q$ and $2q$ spin spiral states, however, we increase the sampling size: we first choose one central spin at the origin and calculate its interactions with its nearest neighbors, and then add the interactions of the six nearest-neighbor spins with their nearest neighbors for a total of 30 bonds. 
Because the $1q$ and $2q$ spin spiral states can have a long period, it is important to accurately include the anisotropy energy as easy-plane anisotropy frustrates the DMI. The anisotropy energy and the Zeeman energy are therefore averaged over a large supercell with $N_{\rm site}$ sites, {\em i.e.,}
\begin{equation}
    E_{\rm Z}/{\rm Spin}=-\frac{1}{N_{\rm sites}}\mathbf{H}_{\rm ext}\cdot\sum_{n_a,n_b}
    {\mathbf S}\left[ \mathbf{q}\cdot(n_a{\mathbf b}_1+n_b{\mathbf b}_2)+\varphi\right],
\end{equation}
with $N_{\rm sites}$ typically $16\times16$ to $25\times25$. 
We directly minimize the energy total energy per spin 
with respect to variational parameters of the $3q$ and spin spiral states to obtain the variational ground state. 

As discussed earlier, a collinear AFM cannot yield a non-zero AHE, and a  co-planar AFM has no contribution to the AHE from the spin chirality. The AHE is directly related to the chirality $\chi$ of the spin structure which we calculate as
\begin{equation}
    \chi=\epsilon_{ijk}{\mathbf S}(\tau_i)\cdot\left[ 
    {\mathbf S}(\tau_j)\times{\mathbf S}(\tau_k)\right],
\end{equation}
where $\epsilon_{\alpha\beta\gamma}$ is the Levi-Civita symbol, repeated indices are summed over, and the sites $i$, $j$, and $k$ form an elementary triangular plaquette. 

We explored the phase space for a range of $B$ between 0.025 and 0.4, and $K$ ranging from 0.025 to 0.1. The resulting phase diagram evolves slowly with varying $B$ and $K$, in particular the dependence on $K$ is rather weak. We will therefore typically discuss results for $B$ in the range of 0.3 to 0.4, with $K=0.05$ or $K=0.1$.

\section{Results and Discussion}\label{sec:results}

For zero DMI coupling $D$, and zero next-nearest neighbor coupling $J_2$, and also for all values of $B$ we have examined, the ground-state is the well-known N\'eel triangular AFM state with the three spins on an elementary triangular plaquette $120^\circ$ apart, and with the out-of-plane anisotropy $K>0$, the spins are co-planar in the $xy$-plane (see Fig.~\ref{fig:1q_2q_ex}). This state is captured by the $1q$ and $2q$ {\em Ans\"atze} but not by the $3q$ {\em Ansatz}, and the $1q$ and $2q$ states correctly yield the ground state. Because the spins are co-planar, the state has a vanishing chirality and therefore vanishing anomalous Hall conductivity. When the interactions $B$, $J_2$, and $D$ are increasing from zero, the spin structure becomes more complicated. For small $J_2$ in the range of 0.1 at $D=0$ to about 0.3 at $D=0.5$, the $1q$ state is always lower in energy than the $2q$ state (the $2q$ variational state collapses to the $1q$ state; we also confirmed numerically that the $1q$ and $2q$ {\em Ans\"atze} are degenerate for small $J_2$).

\begin{figure}
    \centering
    \includegraphics[width=\columnwidth]{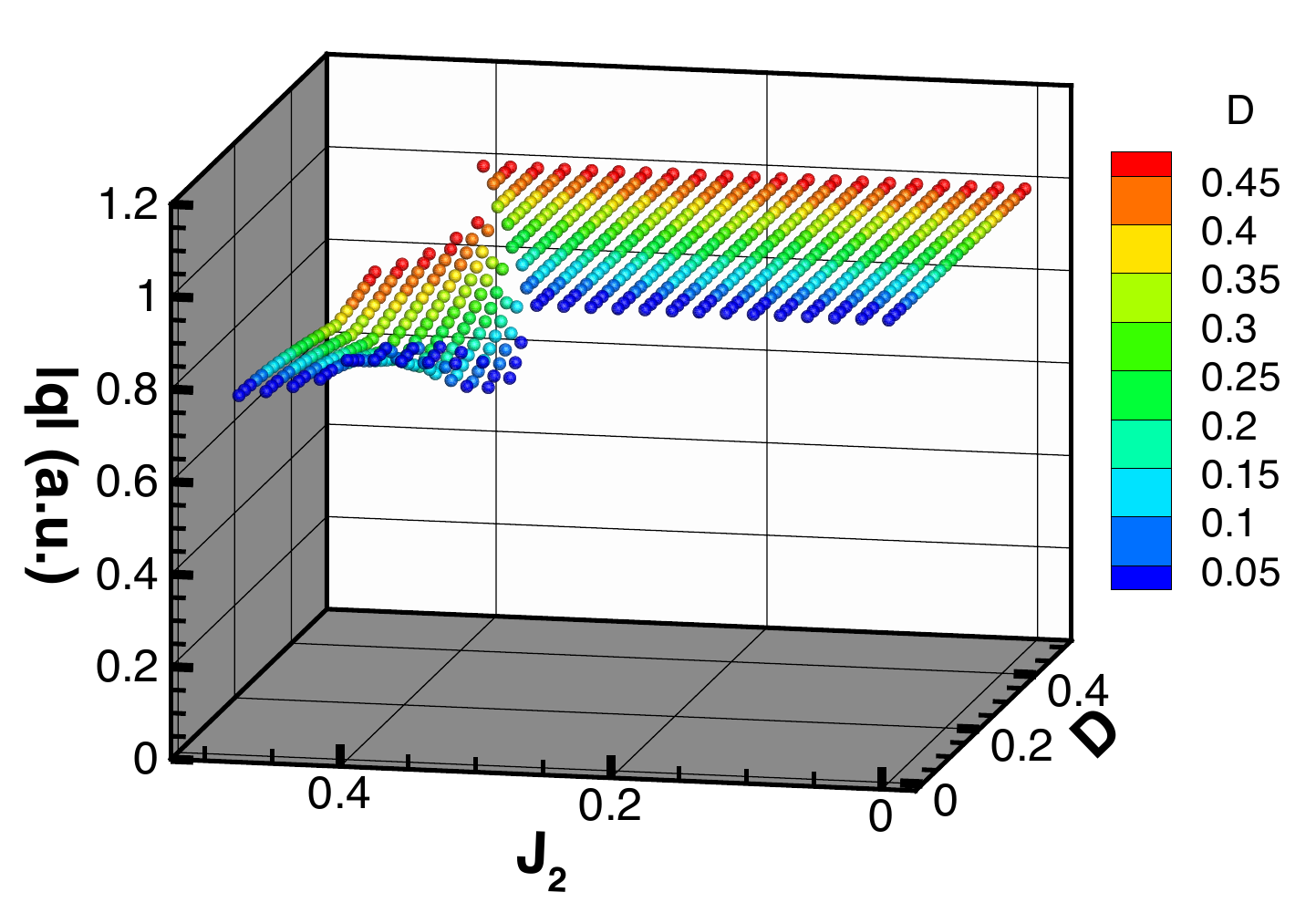}
    \caption{The figure shows the evolution of the norm of $\mathbf{q}$ normalized to the K point in the BZ for the $1q$ state as function of $J_2$ and $D$ for $B=0.3$ and $K=0.1$. The color coding indicates the value of $D$. $|\mathbf {q}|$ is unity as $\mathbf{q}$ is at a K point for small $J_2$ and $D$. For $J_2$ above some critical value, the norm of $\mathbf{q}$ starts to decrease as $\mathbf{q}$ moves from the K points in towards the zone center $\Gamma$. This critical value of $J_2$ increases slightly with $D$.}
    \label{fig:1q_qxqy_v2}
\end{figure}

\begin{figure*}
    \centering
    \includegraphics[width=3.5truein]{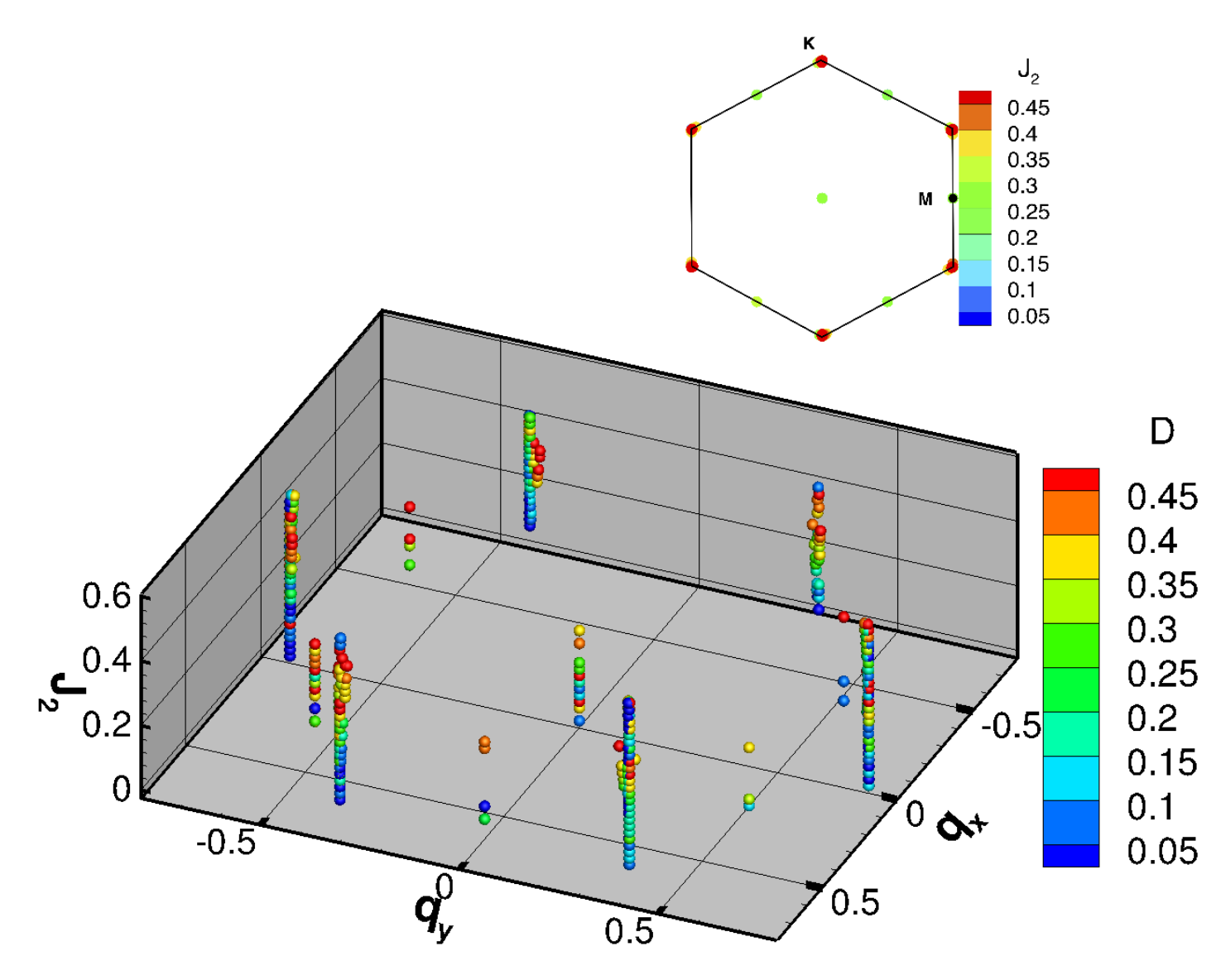}
    \includegraphics[width=3.5truein]{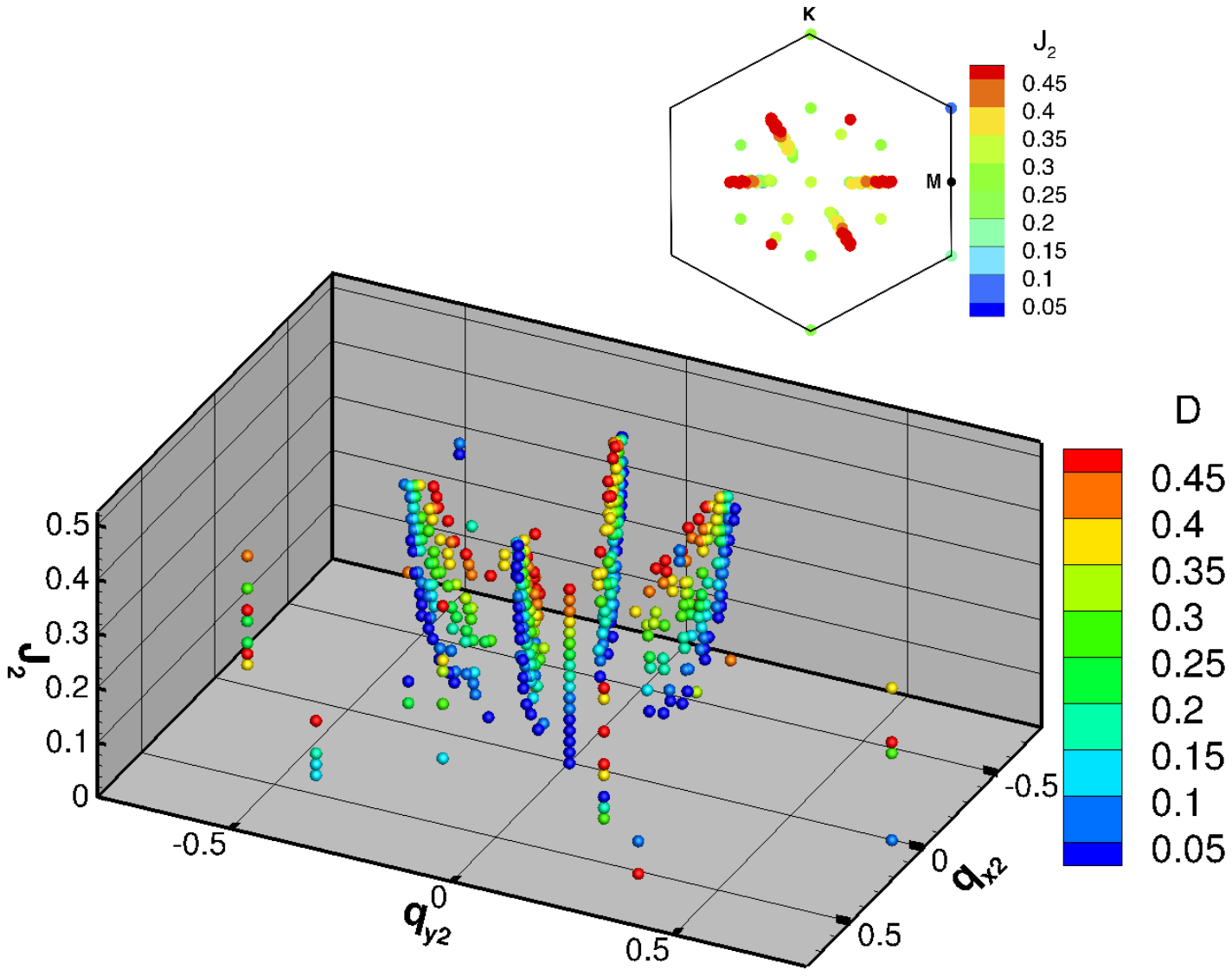}
    \caption{The panels show the evolution of wavevectors $\mathbf{q}$ (left panel) and $\mathbf{q}_2$ (right panel) for the $2q$ state as function of $J_2$ for $B=0.3$ and $K=0.1$. The color coding indicates the value of $D$. The insets show the positions of the wavevectors in the 1st BZ, with the color coding denoting $J_2$. $\mathbf q$ is generally at the K-points but move slightly inwards towards the zone center as $D$ increases, more so as $B$ decreases below $B=0.3$. Occasionally for small values of $J_2$, $J_2\alt0.2$, $\mathbf q$ is at the $\Gamma$ or at M points, in which case $\mathbf {q}_2$ is at the K points or halfways to the K points. $\mathbf {q}_2$ moves towards the M points from $\Gamma$ approximately as the square root of $J_2$ with increasing $J_2$; as $D$ increases, a larger $J_2$ is required to move $\mathbf{q}_2$ from the zone center.} 
    \label{fig:1q_2q_B03}
\end{figure*}

\begin{figure*}
    \centering
    \includegraphics[width=3.5truein]{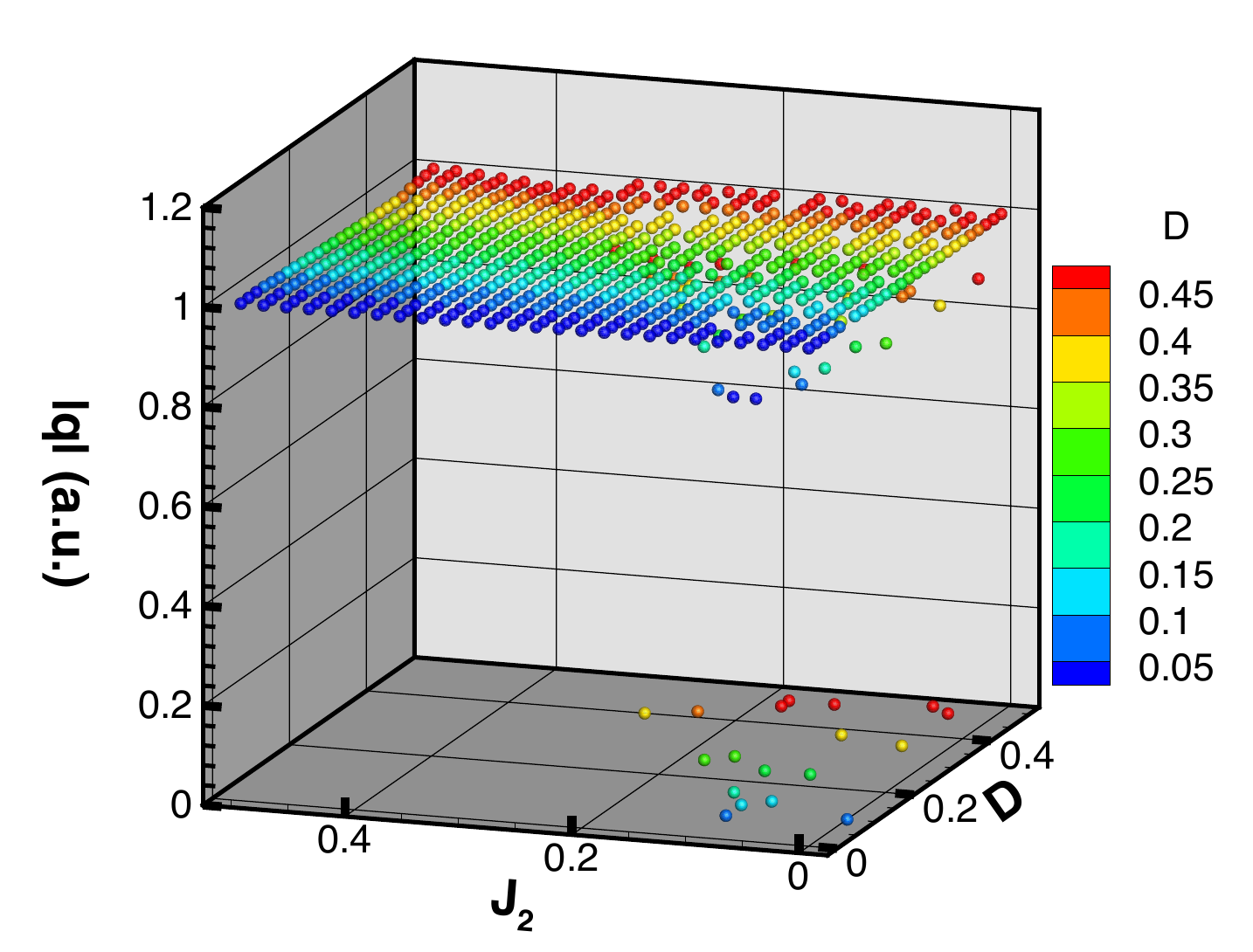}
    \includegraphics[width=3.5truein]{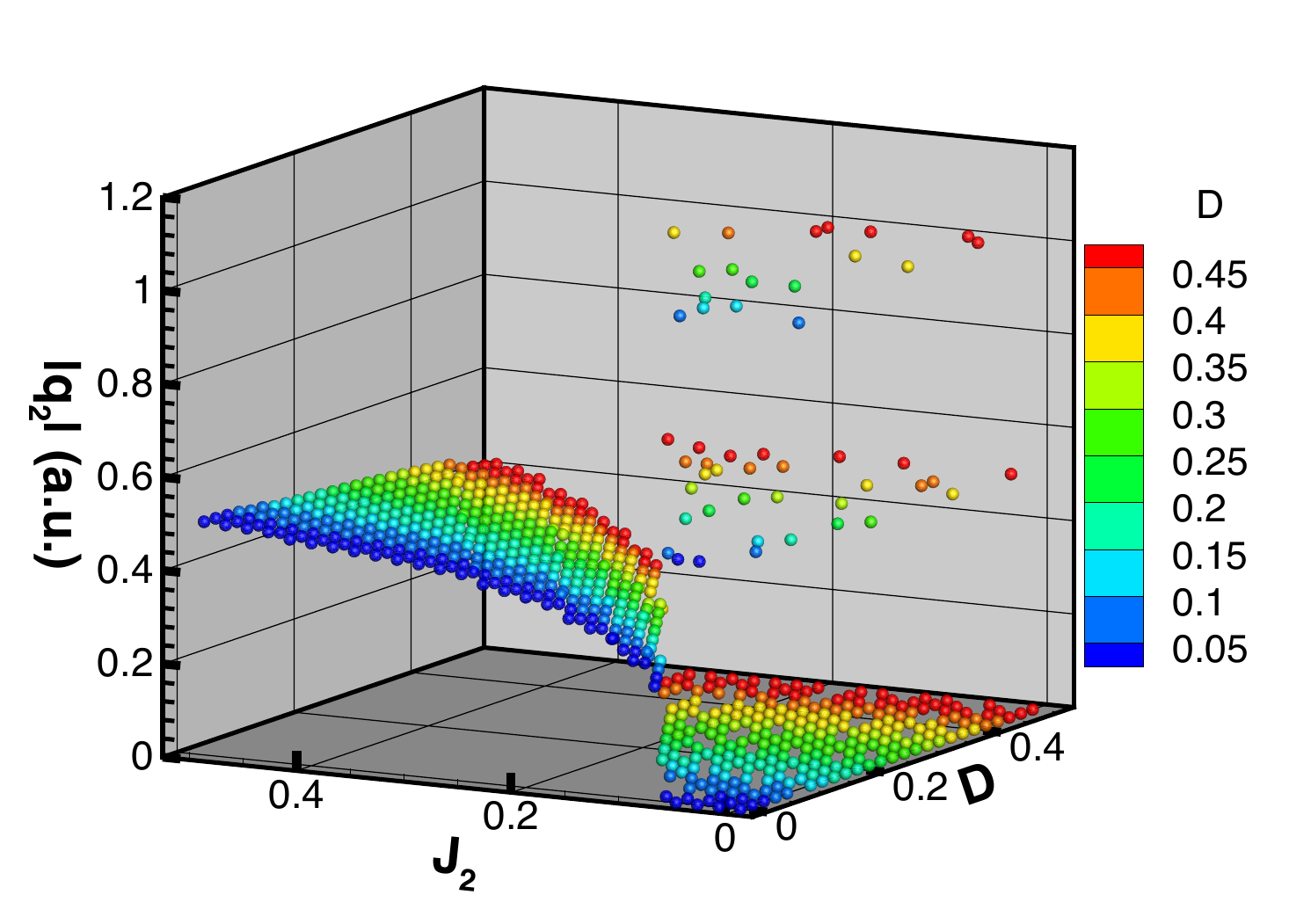}
    \caption{The left (right) right panel shows the evolution of $|\mathbf{q|}$ (left panel) and $|\mathbf{q}_2|$ (right panel) normalized to the K point in the BZ for the $2q$ state as function of $J_2$ and $D$ for $B=0.3$ and $K=0.1$. The color coding indicates the value of $D$. $|\mathbf{q}|$ is unity as $\mathbf{q}$ is at a K point for small $J_2$ and $D$, except for a few scattered points; these are all just other representations of the $1q$ state. $|\mathbf{q}_2|$ is zero for small $J_2$ and $D$, but starts to grow as $J_2$ exceeds a critical value that depends on $D$ and grows approximately linearly with D: when $D=0$, this critical value is about 0.06, and when $D=0.5$, the critical value is about 0.3. The critical value $J_2(D)$ marks the transition from a $1q$ ground state to a $2q$ ground state.}
    \label{fig:2q_qxqy_v2}
\end{figure*}

\begin{figure}
    \centering
    \includegraphics[width=\columnwidth]{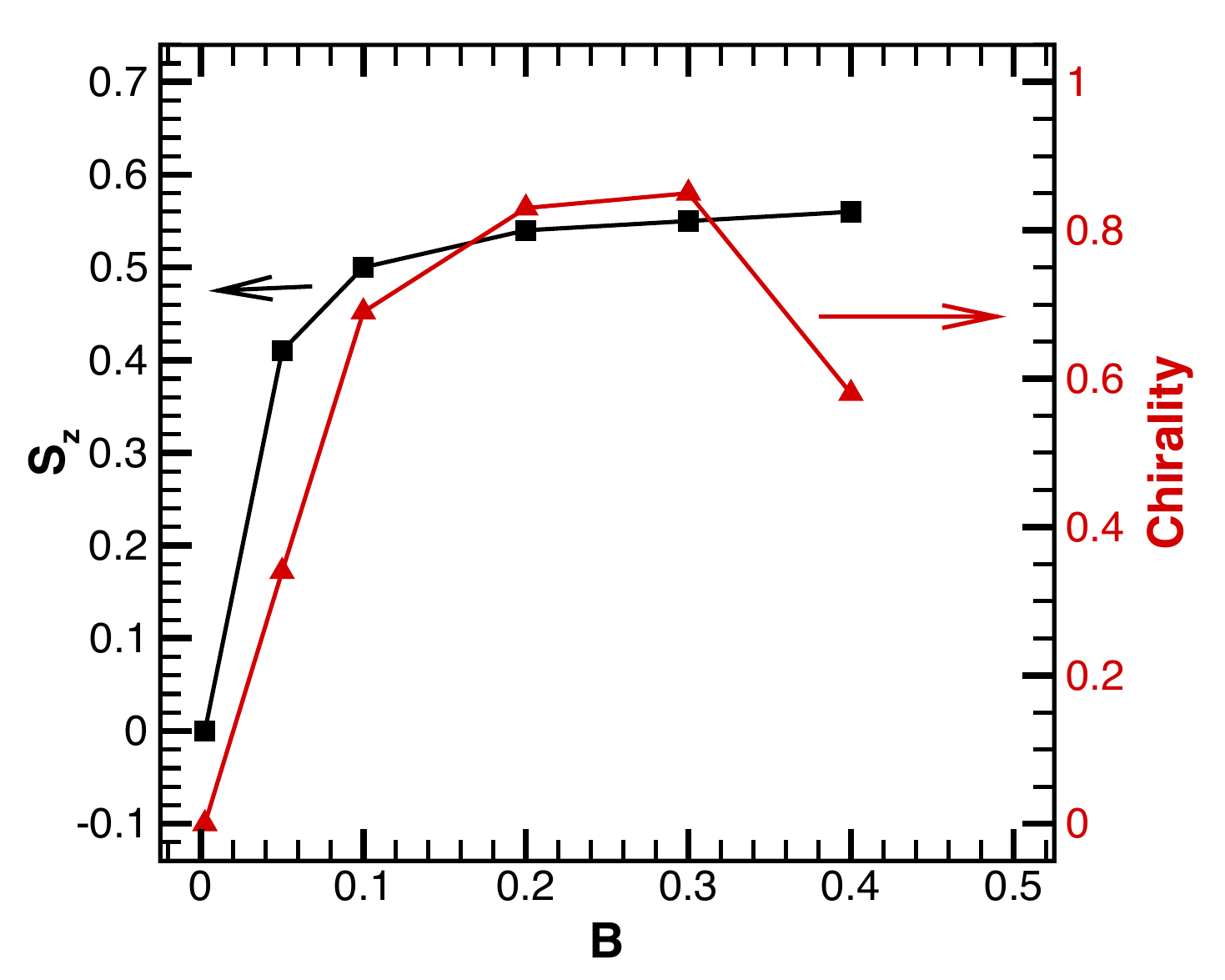}
    \caption{Magnitude of the $z$-components of the spins (black squares and black line) and average chirality (red diamonds and red line) in the $3q$ state as function of $B$ for $K=0.1$ and $J_2=0$.}
    \label{fig:3q_cos_theta}
\end{figure}

Figure~\ref{fig:1q_qxqy_v2} shows $|\mathbf{q}|$ normalized to the K point in the first BZ in the $1q$ state as function of $J_2$ and $D$ for $B=0.3$ and $K=0.1$. For small $J_2$ and $D$, $\mathbf{q}$ falls on the K points and the state is a planar N\'eel state. As $J_2$ increases above some critical value that depends weakly on $D$, $\mathbf{q}$ moves in towards the zone center $\Gamma$ as the interaction parameters try to drive the system towards an incommensurate spiral that is in general non-planar. If $B$ is too small, $B\alt0.3$ (a value that depends very weakly on $K$), the interactions cannot drive $\mathbf{q}$ away from the BZ boundary, and instead of moving in towards $\Gamma$, $\mathbf{q}$ moves on the BZ boundary. 
For the $2q$ state, $\mathbf{q}$ in general falls on the K points on the BZ boundary (see Fig.~\ref{fig:1q_2q_B03}). For small $J_2$, $J_2\alt0.1$, $\mathbf {q}_2$ is at the zone center $\Gamma$ (which makes the state a $1q$ state) but increases approximately as the square-root of $J_2$ with increasing $J_2$ and moves towards the $M$ points, stopping half-ways to the M points; the larger $D$ is, the larger $J_2$ has to be for $\mathbf{q}_2$ to start moving from the zone center. For a few values of $J_2$ and $D$, generally with $J_2\alt0.2$, $\mathbf{q}$ falls at the M points and $\mathbf{q}_2$ on the K points; these particular $2q$ states are in fact another representation of $1q$ states. Figure~\ref{fig:1q_2q_B03} also displays a six-fold symmetry, as we have not folded the obtained values of $\mathbf{q}$ and $\mathbf{q}_2$ back to an irreducible wedge of the 2S BZ.  
Figure~\ref{fig:2q_qxqy_v2} shows the evolution of $|\mathbf{q}|$ and $|\mathbf{q}_2|$ normalized to the K point in the $2q$ state as functions of $J_2$ and $D$. For small $J_2$ and $D$, $\mathbf{q}$ is on a K point and $\mathbf{q}_2=0$, and the $2q$ state is equivalent to the $1q$ (this is also the case for the few scattered points at which $\mathbf{q}$ is at M points or is zero). For some critical value of $J_2$, the norm of $\mathbf{q}_2$ suddenly increases and $\mathbf{q}_2$ starts to move towards M points in the BZ with $|\mathbf{q}_2|$ growing approximately $\sqrt{J_2}$. The critical value of $J_2$ depends on $D$, and is about 0.06 for $D=0$, and about 0.3 for $D=0.5$; this critical value is the transition from a $1q$ ground state to a $2q$ ground state. The critical value is almost independent of $B$ and very weakly dependent on $K$.  
The noncoplanar $3q$ state is stabilized for $B>0$ by $J_2>0$. The dependence on $B$ is stronger than for the spin spiral states, in that the magnitude of the $z$-component of the spins, $S_z$, and the chirality increase rapidly with $B$ for fixed $K$ (see Fig.~\ref{fig:3q_cos_theta}). The dependence on $K$ is weak, except that for very small $B$, $B\alt0.025$, the chirality and $S_z$ components are zero for $K$ too large, $K\agt0.05$.  In contrast, the chirality of the $1q$ spin spiral state is always zero. The net magnetization is zero in the absence of an external field. The energy of the $3q$ state is also independent of $D$. 

\begin{figure}
    \centering
    \includegraphics[width=\columnwidth]{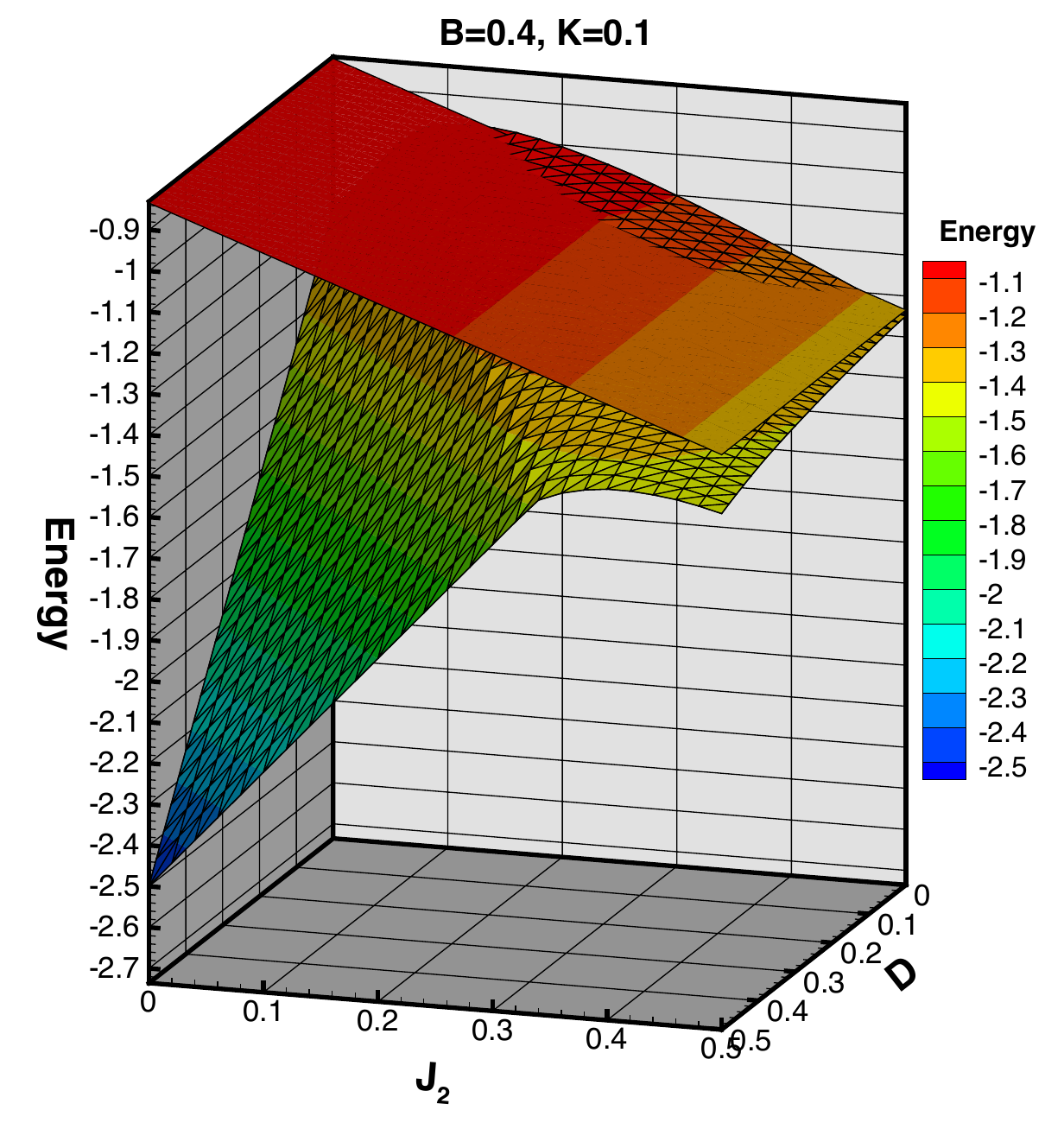}
    \caption{Energy surfaces of the $2q$ state (hatched) and the $3q$ state for $B=0.4$ and $K=0.1$. The $3q$ state has lower energy as $J_2$ increases.}
    \label{fig:energy_surfaces}
\end{figure}

While the $1q$ (or $2q$) state yields the correct ground state for $B=D=J_2=0$, for small but finite $B$ and $J_2$, we would expect the ground state of CoNb$_3$S$_6$ to be the $3q$ state based on electronic structure calculations\cite{hyowon}; these also confirm that this state has a nonzero AHE. This implies that there must be a transition from a spin spiral to a $3q$ state as the interaction parameters are increased. This, in turn, makes it interesting to explore the phase diagram of this system as a transition between $3q$ and spin spiral states could have an immediate observable consequence in the AHE. Figure \ref{fig:energy_surfaces} depicts the energy surfaces of the $2q$ and $3q$ states for $B=0.4$ and $K=0.1$. While the energy for the $3q$ state is independent of $D$ and decreases linearly with increasing $J_2$, the energy surface of the $2q$ state has a local maximum as function of $J_2$ for fixed $D$. As a consequence, the two energy surfaces intersect at large enough $B$, $B\agt0.3$, for small $D$, and the $3q$ state has lower energy for a range of $D$ and $J_2$. 

\begin{figure}
    \centering
    \includegraphics[width=\columnwidth]{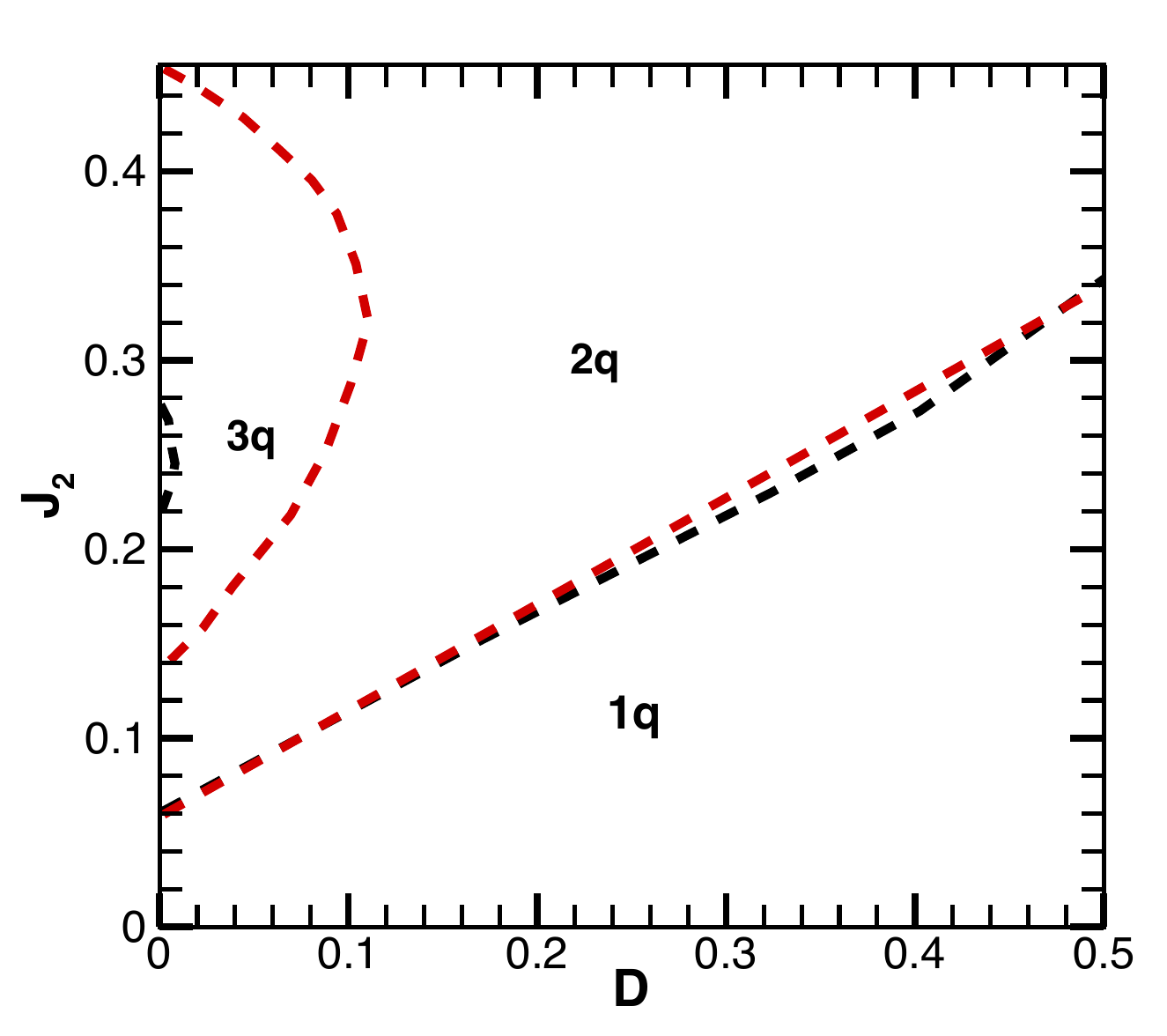}
    \caption{Phase diagram on the $D$-$J_2$ space for $K=0.1$ and for $B=0.3$ (black dashed lines), and $B=0.4$ (red lines). The $1q$ state occupies the phase diagram for small $J_2$. The $3q$ phase emerges near $D=0$ for $J_2$ near 0.25 as $B$ increases.}
    \label{fig:phase_diagrams}
\end{figure}

Figure~\ref{fig:phase_diagrams} shows the phase diagrams in the $D$-$J_2$ space for $K=0.1$ with $B=0.3$ and $B=0.4$. The $3q$ state occupies a region with small $D$ and non-zero $J_2$. For $B=0.3$ this phase is barely visible near $D=0$. As $B$ increases, this region increases in size. For $B=0.4$, the $3q$ state occupies a small strip near $D=0$ for small $J_2\alt0.1$ The $1q$ state is always the ground state for small $J_2$. The phase diagram does not change much as 
$B$ increases from 0.4 to 0.5. Furthermore, the dependence on $K$ is weak; decreasing $K$ by a factor of two from $K=0.1$ to $K=0.05$ only very slightly increases the region of the $3q$ state to larger $D$ and a larger range of $J_2$ by less than 0.04 for $J_2$ and about 0.02 for $D$. 

The transition from a co-planar N\'eel state to a non-coplanar $3q$ state with increasing $J_2$, and then to what appears to be a $2q$ state, was also confirmed by finite-T 3D atomistic simulations using the s-LLG equation\cite{garcia1998langevin}, as detailed earlier in this section. 

The $1q$ state has zero chirality because of symmetry {and is also a planar state with the magnetization in the $ab$-plane in the parameter space that we have examined here;} as stated earlier the $3q$ state has a non-zero chirality. For the optimzed $3q$, state, the chirality (averaged over a unit cell) depends strongly on $B$ for small $B$ (see Fig.~\ref{fig:3q_cos_theta}) but very weakly on the other parameters. The $2q$ state has different chirality properties. The local chirality evaluated over any elementary plaquette is in general non-zero as the spins are in general co-planar. However, the chirality oscillates in magnitude and changes sign in space from one triangular plaquette to the next, and the chirality is zero when averaged over a number of plaquettes. This implies that the AHE arising from spin chirality is zero in the $2q$ state as well as in the $1q$ state with its zero chirality. Therefore, the  $1q$ and $2q$ spin spiral states cannot give rise to a Berry phase and a non-zero AHE from the spin chirality alone, while the $3q$ state can. 

%
The $1q$, $2q$, and $3q$ states all have net zero magnetization along any axis.  The out-of-plane susceptibility for the $1q$, $2q$ and $3q$ states is small and relatively uninteresting, at least for the parameter ranges we have investigated. For fields up to $H_z=0.1$, the average $S_z$ component, $\langle S_z\rangle$, grows linearly by a small amount of up to about $0.01$. 
A small out-of-plane susceptibility is consistent with the results for CoNb$_3$S$_6$ by Ghimire {\em et al.}\cite{ghimire2018large}.

\section{Conclusions and summary}\label{sec:discussion}
We have here proposed and analyzed a model for the in-plane magnetic interactions in the family of triangular AFMs in transition-metal intercalated dichalcogenides (TM)Nb$_3$S$_6$. The model allows us to search for three general classes of magnetic ground states, $1q$ and $2q$ spin spiral states, and a $3q$ state with four spins per unit cell. For small in-plane next-nearest neighbor interactions $J_2\alt0.1$, the $1q$ spin spiral state is the ground state, but with increasing $J_2$, the system transitions to a $2q$ state which generally is non-coplanar. A non-zero $J_2$ lead to a $3q$ ground state stabilized by a non-zero $B$. For $B\agt0.3$ the non-coplanar $3q$ state emerges as the ground state for a range of the Dzhyaloshinskii-Moriya interaction $D\ge0$. The $1q$ and $2q$ states have vanishing chirality $\chi={\mathbf S}_1\cdot\left[ {\mathbf S}_2\times{\mathbf S}_3\right]$ evaluated over the three spins in an elementary triangular plaquette and averaged over many plaquettes, and so the spin chirality will not contribute to an AHE signal for these states. The non-coplanar $3q$ state has a non-zero chirality. In a 3D system, this gives rise to a non-zero anomalous Hall effect provided the stacked 2D layers have the same chirality; a large non-zero AHE is consistent with measurements\cite{ghimire2018large,tenasini2020giant} on CoNb$_3$S$_6$ and electronic structure calculations that include spin-orbit interactions\cite{hyowon}. While our model suggests that the $3q$ structure in CoNb$_3$S$_6$ can give rise to an observed AHE, we cannot make any quantitative predictions about the magnitude of the quantum Hall conductivity. This is because CoNb$_3$S$_6$ is a metal with the Co-hybridized bands crossing the Fermi level\cite{hyowon}. Therefore, the actual values of the anomalous Hall conductivity depend sensitively on the details of the electronic structure and is beyond the scope of this work.   
However, increasing $D$ drives the system to non-chiral $2q$ or $1q$ states, as the DMI with its vector along the $c$ axis favors planar spins. Furthermore, too small $B$ will not be able to stabilize the $3q$ state. Therefore, the observed AHE\cite{ghimire2018large,tenasini2020giant} puts constraints on $D$ and $B$: $B$ must be approximately greater than 0.2, and $D$ must be smaller than approximately 0.1 in order to drive a drive the system to a $3q$ with non-zero chirality.

The sensitivity of the ground state to interaction parameters opens the intriguing possibility of inducing a transition between the $3q$ state and the spin spiral states by, for example, bi-axial in-plane strain. Another potential mechanism is substitutional doping, e.g.\ Mn for Co. The MnNb$_3$S$_6$ ground state is a planar ferromagnet\cite{hyowon}, presumably because of stronger double-exchange, and Mn is much more likely to occupy Co sites than interstitial or Nb/S sites. Doping may change the magnetic interactions (and also the electron filling) and induce a transition, although there is a small possibility that doping may lead to more complicated interactions not considered here. Such a transition from $3q$ to $2q$ or $1q$ states induced by strain or doping should have a clear signature in the magnetotransport properties.

\begin{acknowledgments}
We gratefully acknowledge insightful conversations with I. Martin and J.F. Mitchell. OH and HP acknowledge funding from the US Department of Energy, Office of Science, Basic Energy Sciences Division of Materials Sciences and Engineering.
RAH received funding from the European Research Council (ERC) under the European Union’s Horizon 2020 research and innovation programme (grant agreement No 882340). We gratefully acknowledge the computing resources provided on Bebop, Swing, and Blues, high-performance computing clusters operated by the Laboratory Computing Resource Center at Argonne National Laboratory.
\end{acknowledgments}

\bibliography{CNS}
\end{document}